\documentclass[pre,twocolumn,amsmath,amssymb,floatfix,superscriptaddress,nopacs]{revtex4}

\usepackage{graphicx}
\usepackage{latexsym}
\usepackage{amsmath}
\usepackage{amssymb}
\usepackage{amsfonts}
\usepackage{bm}
\usepackage{color}

\newcommand{\la}{\left<}
\newcommand{\ra}{\right>}

\newcommand{\nvecl}{\underline{n}_l}
\newcommand{\pvec}{\ensuremath{\underline{p}}}

\newcommand{\rvec}{\ensuremath{\underline{r}}}
\newcommand{\rvecl}{\ensuremath{\underline{r}_l}}

\newcommand{\ddiff}{\ensuremath{\mathrm{d}}}
\newcommand{\tp}{\ensuremath{t^{\prime}}}

\newcommand{\rx}{r_x}
\newcommand{\ry}{r_y}
\newcommand{\px}{p_x}
\newcommand{\py}{p_y}
\newcommand{\pix}{p_{ix}}
\newcommand{\piy}{p_{iy}}

\newcommand{\rl}{r_{l}}
\newcommand{\nlx}{n_{lx}}
\newcommand{\nly}{n_{ly}}

\newcommand{\kB}{k_\mathrm{B}}

\newcommand{\Ehat}{\hat{E}}
\newcommand{\ehat}{\hat{e}}
\newcommand{\Eidhat}{\hat{E}_\mathrm{id}}
\newcommand{\Eexhat}{\hat{E}_\mathrm{ex}}

\newcommand{\sxy}{\ensuremath{\sigma}}
\newcommand{\tauhat}{\ensuremath{\hat{\sigma}}}
\newcommand{\tauidhat}{\ensuremath{\hat{\sigma}_\mathrm{id}}}
\newcommand{\tauexhat}{\ensuremath{\hat{\sigma}_\mathrm{ex}}}

\newcommand{\muA}{\ensuremath{\mu_\mathrm{A}}}

\newcommand{\muAsurf}{\ensuremath{\mu_\mathrm{As}}}
\newcommand{\muAhat}{\ensuremath{\hat{\mu}_\mathrm{A}}}
\newcommand{\muAidhat}{\ensuremath{\hat{\mu}_\mathrm{id}}}
\newcommand{\muAexhat}{\ensuremath{\hat{\mu}_\mathrm{ex}}}

\newcommand{\muF}{\ensuremath{\mu_\mathrm{F}}}

\newcommand{\muFbulk}{\ensuremath{\mu_\mathrm{F0}}}
\newcommand{\muFsurf}{\ensuremath{\mu_\mathrm{Fs}}}
\newcommand{\muFone}{\ensuremath{\mu_1}}
\newcommand{\muFtwo}{\ensuremath{\mu_0}}

\newcommand{\Geq}{\mu_\mathrm{eq}}
\newcommand{\GF}{\mu}

\newcommand{\GFglass}{\mbox{$\mu_\mathrm{g}$}}
\newcommand{\GFplat}{\mbox{$\mu_\mathrm{p}$}}
\newcommand{\dGF}{\delta \mu}
\newcommand{\dGt}{\delta G(t)}

\newcommand{\Gstor}{\mbox{$G^{\prime}(\omega)$}}
\newcommand{\GstorT}{\mbox{$G^{\prime}(\omega,T)$}}
\newcommand{\Gloss}{\mbox{$G^{\prime\prime}(\omega)$}}
\newcommand{\GlossT}{\mbox{$G^{\prime\prime}(\omega,T)$}}

\newcommand{\Lz}{\mbox{$L_{\rm z}$}}
\newcommand{\ULJ}{U_\mathrm{LJ}}
\newcommand{\ULJshift}{U_\mathrm{LJ,trunc}}
\newcommand{\epsLJ}{\epsilon_\mathrm{LJ}}
\newcommand{\sigLJ}{\sigma_\mathrm{LJ}}
\newcommand{\Ubond}{U_\mathrm{bond}}
\newcommand{\rmin}{r_\mathrm{min}}
\newcommand{\rcut}{r_\mathrm{cut}}
\newcommand{\lbond}{l_\mathrm{bond}}
\newcommand{\kbond}{k_\mathrm{bond}}
\newcommand{\dtMD}{\delta t_\mathrm{MD}}
\newcommand{\ttemp}{\Delta t_\mathrm{temp}}
\newcommand{\tsampmax}{\Delta t_\mathrm{max}}

\newcommand{\tsamp}{\Delta t}

\newcommand{\Tglass}{\mbox{$T_{\rm g}$}}
\newcommand{\Hglass}{\mbox{$H_{\rm g}$}}

\newcommand{\tauterm}{\tau}

\newcommand{\Rend}{\mbox{$R_{\rm e}$}}
\newcommand{\Rgyr}{\mbox{$R_{\rm g}$}}

\newcommand{\Abulk}{\mbox{${\cal A}_0$}}
\newcommand{\Asurf}{\mbox{${\cal A}_s$}}
\newcommand{\ahat}{\hat{a}}
\newcommand{\Acal}{\mbox{$\cal A$}}
\newcommand{\Tver}{\ensuremath{{\cal P}_{\tsamp}}}

\begin{document}

\title{Shear-stress relaxation in free-standing polymer films}

\author{G. George}
\affiliation{Institut Charles Sadron, Universit\'e de Strasbourg \& CNRS, 23 rue du Loess, 67034 Strasbourg Cedex, France}
\author{I. Kriuchevskyi}
\affiliation{LAMCOS, INSA, 27 av. Jean Capelle, 69621 Villeurbanne Cedex, France}
\author{H. Meyer}
\affiliation{Institut Charles Sadron, Universit\'e de Strasbourg \& CNRS, 23 rue du Loess, 67034 Strasbourg Cedex, France}
\author{J. Baschnagel}
\affiliation{Institut Charles Sadron, Universit\'e de Strasbourg \& CNRS, 23 rue du Loess, 67034 Strasbourg Cedex, France}
\author{J.P.~Wittmer}
\email{joachim.wittmer@ics-cnrs.unistra.fr}
\affiliation{Institut Charles Sadron, Universit\'e de Strasbourg \& CNRS, 23 rue du Loess, 67034 Strasbourg Cedex, France}

\begin{abstract}
Using molecular dynamics simulation of a polymer glass model we investigate free-standing 
polymer films focusing on the in-plane shear modulus $\GF$, 
defined by means of the stress-fluctuation formula, as a function of temperature $T$, 
film thickness $H$ (tuned by means of the lateral box size $L$) and sampling time $\tsamp$.
Various observables 
are seen to vary linearly with $1/H$ demonstrating thus the (to leading order) 
linear superposition of bulk and surface properties.
Confirming the time-translational invariance of our systems, $\GF(\tsamp)$ is shown to be
numerically equivalent to a second integral over the shear-stress relaxation modulus $G(t)$. 
It is thus a natural smoothing function statistically better behaved as $G(t)$. 
As shown from the standard deviations $\delta \GF$ and $\delta G$, this is especially 
important for large times and for temperatures around the glass transition.
$\GF$ and $G$ are found to decrease continuously with $T$ and a jump-singularity is not observed.
Using the Einstein-Helfand relation for $\GF(\tsamp)$ and the successful time-temperature superposition    
scaling of $\GF(\tsamp)$ and $G(t)$ the shear viscosity $\eta(T)$ can be estimated for a broad range of 
temperatures.
\end{abstract}
\date{\today}
\maketitle

\section{Introduction}
\label{sec_intro}

\begin{figure}[t]
\centerline{\resizebox{0.9\columnwidth}{!}{\includegraphics*{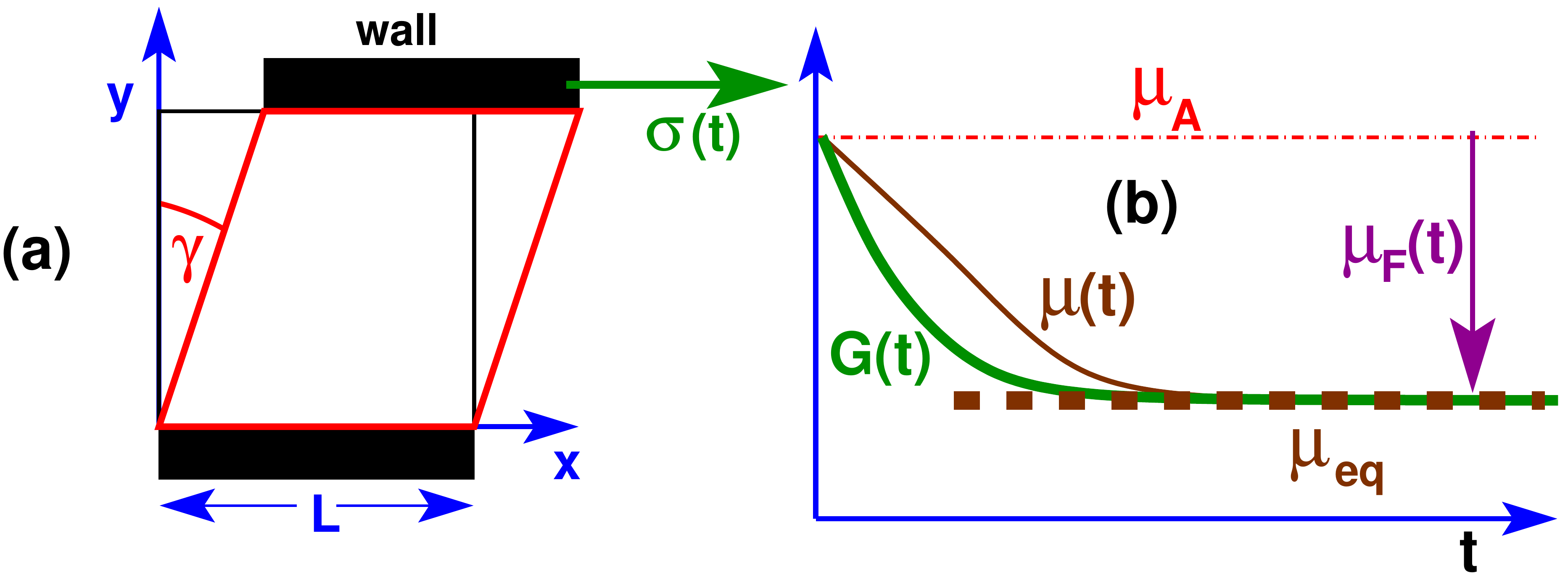}}}
\caption{Some notations:
{\bf (a)}
Simple shear with  $\gamma$ being the strain increment imposed at $t=0$
and $\sxy(t)$ the measured shear stress increment as a function of time $t$.
{\bf (b)}
Shear-stress relaxation modulus $G(t)$ (dash-dotted line) and generalized shear modulus 
$\GF(t)=\muA-\muF(t)$ (thin solid line).
The {\em affine} shear modulus $\muA=G(t=0)=\GF(t=0)$ is indicated by the dash-dotted line,
the thermodynamic long-time limit $\Geq$ for $G(t)$ and $\GF(t)$ by the bold dashed line.
}
\label{fig_sketchSFF}
\end{figure}

\subsection{Generalized shear modulus $\GF(\tsamp)$}
\label{intro_back}

An important mechanical property characterizing elastic solids or more general viscoelastic bodies 
is the thermodynamic equilibrium shear modulus $\Geq$ \cite{FerryBook,RubinsteinBook}. 
(We remind that $\Geq=0$ for simple or complex liquids.)
As sketched in Fig.~\ref{fig_sketchSFF}, $\Geq$ is the long-time limit of the 
shear-stress relaxation modulus $G(t)$, i.e. the ratio of the measured shear stress $\sxy(t)$ and 
the imposed (infinitesimal) simple shear strain $\gamma$. 
Instead of using a tedious out-of-equilibrium simulation tilting the simulation box
as shown in panel (a), the shear modulus may be conveniently obtained numerically
using equilibrium time series of the instantaneous shear stress $\tauhat$ and the 
instantaneous affine shear modulus $\muAhat$ as defined in Appendix~\ref{app_tauAmuA}.
This is done by means of the well-known stress-fluctuation formula
\cite{Hoover69,Barrat88,Lutsko88,SBM11,XWP12,WXP13,WXB15,WXB16,WKC16,LXW16,ivan17c,ivan18a}
\begin{equation}
\GF(\tsamp) \equiv \muA - \muF(\tsamp)
\mbox{ with } \GF(\tsamp) \to \Geq
\label{eq_SFF}
\end{equation}
in the limit of a sufficiently large sampling time $\tsamp$ of the computer experiment
\cite{foot_samplingtime}.
As sketched in panel (b) of Fig.~\ref{fig_sketchSFF}, the ``affine shear modulus" $\muA$ 
describes the elastic response assuming an infinitesimal canonical affine strain 
(Appendix~\ref{app_afftrans}) of all parts of the body under the macroscopic 
simple shear constraint. 
Correcting the resulting overestimation of the modulus, the non-affine contribution $\muF(\tsamp)$ 
measures the fluctuations of $\tauhat$. (For details see Sec.~\ref{res_SFF}.)
The indicated $\tsamp$-dependences naturally arise since the averages for $\muA$ and $\muF$ are 
commonly and most conveniently done by first ``time-averaging" over time windows of length $\tsamp$ 
of the stored data entries of a given configuration and only in a second step by ``ensemble-averaging" over 
completely independent configurations (Appendix~\ref{app_data}). 
Assuming the time-translational invariance of the time series 
it can be demonstrated (Appendix~\ref{app_fluctu}) 
that the $\tsamp$-dependence can be traced back to the stationarity relation 
\cite{WXB15,WKC16,ivan18a}
\begin{equation}
\GF(\tsamp) = \frac{2}{\tsamp^2} \int_0^{\tsamp}\ddiff t \int_0^t \ddiff \tp \ G(\tp).
\label{eq_muGt} 
\end{equation}
Being a second integral over $G(t)$, $\GF(\tsamp)$ is a convenient smoothing function with 
in general much better statistical properties as $G(t)$.
The historically thermodynamically rooted stress-fluctuation formula, Eq.~(\ref{eq_SFF}),
takes due to Eq.~(\ref{eq_muGt}) the meaning of a generalized quasi-static modulus also containing 
information about dissipation processes associated with the reorganization of the particle contact network.
This has been extensively tested for self-assembled transient networks \cite{WKC16}.

\subsection{Shear modulus of glass-forming systems}
\label{intro_backpoly}
The (thermodynamically well-defined) shear modulus $\Geq(T)$ of crystalline solids is known 
to vanish discontinuously at the melting point with increasing temperature $T$ \cite{Barrat88,LXW16}.
This begs the question of whether $\Geq$ or a natural generalization, such as $\GF(\tsamp)$
describing also stationary out-of-equilibrium systems and general viscoelastic bodies,
behave similarly for amorphous glass-forming colloids or polymers at their glass transition temperature $\Tglass$ 
\cite{Barrat88,Szamel11,Ikeda12,Yoshino12,Yoshino14,Klix12,Klix15,WXP13,ZT13,SBM11,LXW16,ivan17c,ivan18a}.
Qualitative different theoretical \cite{Ikeda12,Yoshino12,Yoshino14,ZT13}, 
experimental \cite{Klix12,Klix15} or numerical \cite{Barrat88,WXP13,ivan17c,ivan18a} 
findings have been put forward suggesting either a {\em discontinuous jump}  
\cite{Szamel11,Ikeda12,Yoshino14,Klix12,Klix15} 
or a {\em continuous} transition 
\cite{Barrat88,Yoshino12,ZT13,WXP13,LXW16,ivan17c,ivan18a}.
Following the pioneering work of Barrat {\em et al} \cite{Barrat88} various numerical
studies have used the stress-fluctuation formula, Eq.~(\ref{eq_SFF}), as the main diagnostic
tool to characterize the shear strain response \cite{Barrat88,WXP13,SBM11,LXW16,ivan17c,ivan18a}.
Using molecular dynamics (MD) simulation \cite{AllenTildesleyBook,FrenkelSmitBook} 
of a coarse-grained bead-spring model \cite{LAMMPS,SBM11,ivan17c,ivan18a} 
we have recently investigated $\GF(\tsamp)$ and $G(t)$ for three-dimensional (3D)
polymer melts \cite{ivan17c,ivan18a}. The most important findings are that
\begin{itemize}
\item
the stationarity relation Eq.~(\ref{eq_muGt}) holds for all temperatures,
i.e. the expectation values of $\GF(\tsamp)$ and $G(t)$ are numerically equivalent;
\item
this is not the case for their standard deviations $\delta \GF$ and $\delta G$
for which $\delta \GF(T) \ll \delta G(T)$ holds;
\item
if taken at the same (sampling) time,
$\GF(T)$ and $G(T)$ are found to decrease continuously with $T$; 
\item
$\delta \GF(T)$ and $\delta G(T)$ are non-monotonic with strong peaks slightly below $\Tglass$.
Theoretical calculations for the expectation values of an ensemble of independent configurations
are thus largely irrelevant for predicting the behavior of one configuration.
\end{itemize}

\begin{figure}[t]
\centerline{\resizebox{0.9\columnwidth}{!}{\includegraphics*{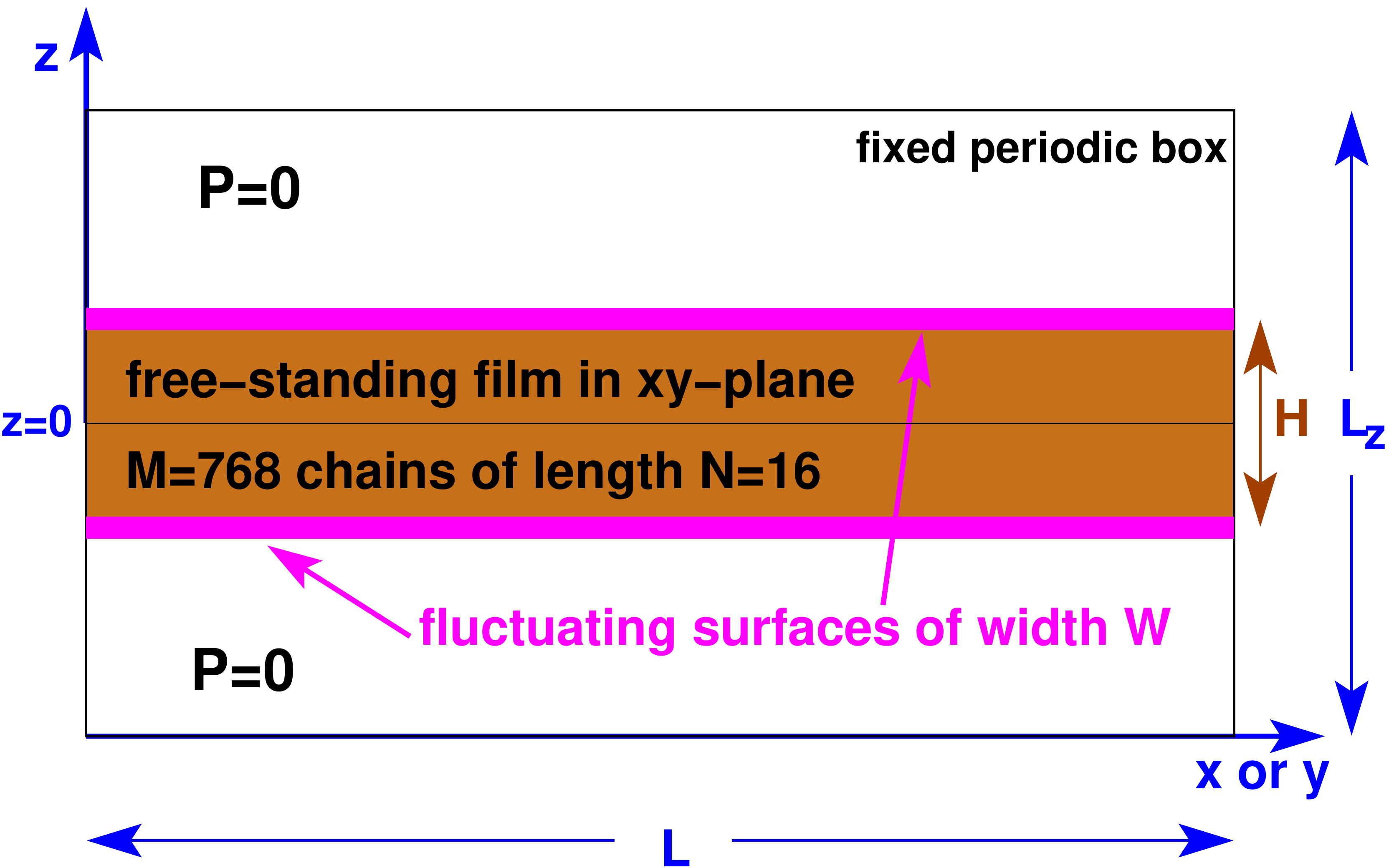}}}
\caption{
We study free-standing polymer films with $M=768$ chains of length $N=16$ monomers confined 
in periodic boxes with $L$ being the imposed lateral box size in both $x$ and $y$ directions. 
The film thickness $H \sim 1/L^2$ (to leading order) is operationally defined 
using the Gibbs dividing surface.
}
\label{fig_sketchSetup}
\end{figure}

\subsection{Aim of present study}
\label{intro_aim}

As sketched in Fig.~\ref{fig_sketchSetup}, the present study extends our previous work 
to free-standing polymer films of finite thickness $H$ tuned by means of the imposed 
lateral box size $L$.
It is well known that the confinement of polymers to thin films can dramatically change
their physical properties 
\cite{KTT95,Dalnoki96,Forrest00,Dalnoki01,Dalnoki01b,Dalnoki12,Dalnoki14,Napolitano07,McKenna05,McKenna05rev,McKenna08,McKenna17,McKenna17b,EllisonTorkelson:NatureMaterials2003,BodiguelFretigny:EPJE2006,YangEtal:Science2010,Roth11,LamTsui:PRE2013,deGennes00,Herming02,Douglas14,Schweizer17,MerabiaEtal:EPJE2004,DequidtEtal:PolymerGlasses2016,MilnerLipson:Macro2010,Varnik00,Varnik02,Peter06,Peter07,Solar12,Pablo00,Pablo02,Pablo02b,Pablo03c,Pablo05b,Riggleman13,LangSimmons:Macro2013,LangEtal:ACSMacroLett2014,MangalaraEtal:JCP2017,ChowdhuryEtal:JPCL2017,Vogt18}. 
Substantial efforts have been made experimentally 
\cite{KTT95,Dalnoki96,Forrest00,Dalnoki01,Dalnoki12,Dalnoki14},
numerically \cite{Pablo00,Pablo02,Pablo02b,Varnik02,Peter06,LangSimmons:Macro2013,MangalaraEtal:JCP2017,ChowdhuryEtal:JPCL2017} 
and theoretically 
\cite{deGennes00,Herming02,Douglas14,Schweizer17,MerabiaEtal:EPJE2004,DequidtEtal:PolymerGlasses2016,MilnerLipson:Macro2010}
to describe the glass transition temperature showing as a general trend that free surfaces lead to 
a decrease of $\Tglass$ \cite{Vogt18}. Despite of their technological importance mechanical and 
rheological properties have been much less studied experimentally 
\cite{BodiguelFretigny:EPJE2006,McKenna05,McKenna08,McKenna17,McKenna17a,Vogt18}. 
(One reason is that much smaller and more precise load cells are required
due to the tiny loads needed to deform the films \cite{Vogt18}.)
Perhaps as a consequence, only a small number of numerical studies exist at present 
focusing on the mechanical properties of films 
\cite{Pablo02b,Pablo03c,Pablo05b,Solar12,Riggleman13,LangEtal:ACSMacroLett2014,ChowdhuryEtal:JPCL2017}
and related amorphous polymer nanostructures \cite{Pablo02b}.
Attempting to fill this gap and using the same coarse-grained numerical model as in 
Refs.~\cite{SBM11,ivan17c,ivan18a}, we focus here on the in-plane shear-stresses,
their fluctuations and relaxation dynamics. 
At variance to real experiments \cite{BodiguelFretigny:EPJE2006,McKenna05,McKenna08,McKenna17,McKenna17a}, 
we use again as the main diagnostic tool the first time-averaged 
and then ensemble-averaged generalized shear modulus $\GF$ and its various contributions as defined by the 
stress-fluctuation formula, Eq.~(\ref{eq_SFF}).
Only total film properties will be discussed for clarity, 
their $z$-resolved contributions will be given elsewhere.

\begin{figure}[t]
\centerline{\resizebox{1.0\columnwidth}{!}{\includegraphics*{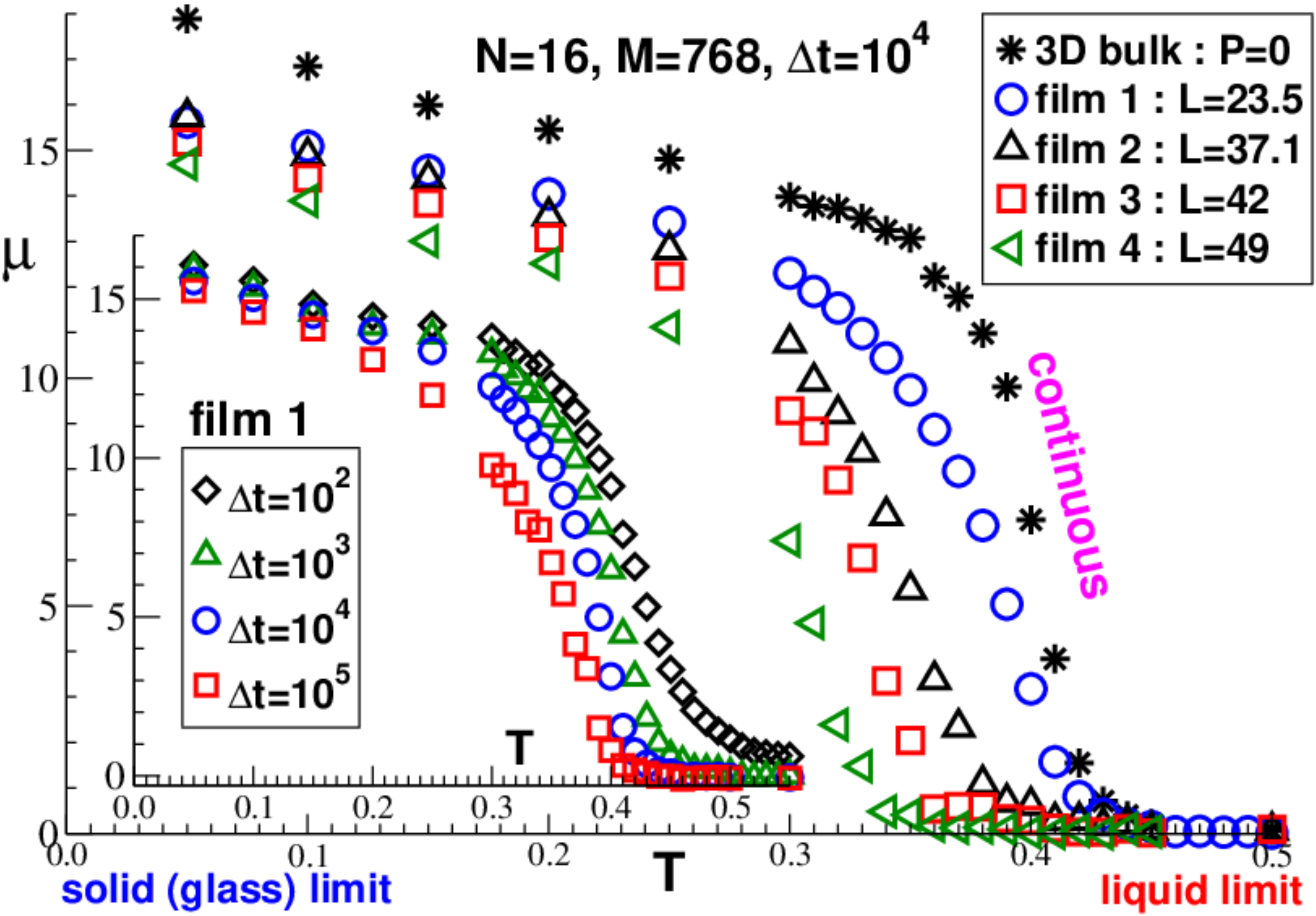}}}
\caption{Shear modulus $\GF$ as calculated by means of Eq.~(\ref{eq_SFF}). 
$\GF(T)$ decays continuously in all cases considered.
Main panel:
Data obtained at a sampling time $\tsamp=10^4$ for three-dimensional bulks (stars)
and films of different lateral box lengths $L$. 
Inset: $\GF(T)$ for film 1 comparing different $\tsamp$.
}
\label{fig_key}
\end{figure}

\subsection{Some key findings}
\label{intro_key}
Summarizing several points made in this paper we present $\GF(T)$ in Fig.~\ref{fig_key} 
for different systems (main panel) and sampling times $\tsamp$ (inset).
As explained in Appendix~\ref{app_Hamiltonian}, Lennard-Jones (LJ) units \cite{AllenTildesleyBook} 
are used here as everywhere in this work.
Confirming our recent work on 3D melts, $\GF(T)$ is observed to decay {\em continuously} in all cases.
Also, as emphasized in the inset, $\GF$ systematically depends on $\tsamp$.  
In addition it is seen in the main panel that $\GF$ becomes 
finite at lower temperatures for thinner films (larger $L$).
We corroborate these findings in the remainder of this paper.
Importantly, Eq.~(\ref{eq_muGt}) will be demonstrated to hold also for polymer films and 
$\GF(\tsamp)$ is thus a natural smoothing function with much better statistics as $G(t)$. 
As shown from the standard deviations $\delta \GF$ and $\delta G$, this is especially
important for large times and for temperatures around the glass transition.
Using the successful time-temperature superposition (TTS) of $\GF(\tsamp)$ and $G(t)$ 
it will be shown that the shear viscosity $\eta(T)$ can be estimated for a broad range of temperatures.

Many intensive properties $\Acal$, such as $\Tglass$, $\muA$ or $\muF$, 
will be seen to depend linearly on the inverse film thickness $H$. 
This is expected for small chains (having a gyration radius $\Rgyr \ll H$)
assuming as the simplest phenomenological description the linear superposition 
\begin{eqnarray}
\Acal & \approx & \frac{1}{H} \left[\Abulk \ (H-W) + \Asurf \ W \right] \nonumber \\
        & = & \Abulk \left[ 1- \frac{(1-\Asurf/\Abulk) W}{H} \right]
\label{eq_Acalsurfeff}
\end{eqnarray} 
of a bulk term $\Abulk$ with a weight $H-W \approx H$ and a surface term $\Asurf$ with a weight 
proportional to the surface width $W \ll H$ \cite{foot_otherH}. 
Even more generally, $\Acal$ may be written as an average
(possibly non-trivially weighted \cite{MangalaraEtal:JCP2017}) 
over $z$-dependent contributions $\Acal(z)$ as done, e.g., 
for the glass transition temperature $\Tglass$ 
\cite{Schweizer17,Peter07,MangalaraEtal:JCP2017} or 
the storage and loss moduli $\Gstor$ and $\Gloss$ \cite{Pablo05b}.
The claimed $1/H$-correction, Eq.~(\ref{eq_Acalsurfeff}), has merely the advantage to be based 
on a simple and transparent idea.
It may be seen as the leading contribution of a more general $1/H$-expansion \cite{foot_otherH}.
We remind that other $H$-dependences have been suggested \cite{deGennes00,Herming02,Schweizer17}
and fitted with some success \cite{Pablo00,Varnik02,Peter06,Peter07}.

\subsection{Outline}
\label{intro_outline}
The different configuration ensembles are characterized in Sec.~\ref{sec_algo} 
before we present our numerical results in Sec.~\ref{sec_res}. 
We start with the characterization of the film thickness $H$ and 
the glass transition temperature $\Tglass$ (Sec.~\ref{res_density}) and 
discuss then the affine and non-affine contributions $\muA$ and $\muF$
to the shear modulus $\GF$ (Sec.~\ref{res_SFF}).
We turn in Sec.~\ref{res_tsamp} to the $\tsamp$-dependence of time-preaveraged fluctuations
and demonstrate that the stationarity relation Eq.~(\ref{eq_muGt}) holds for films.
Using the Einstein-Helfand relation \cite{AllenTildesleyBook,ivan18a} 
we compute in Sec.~\ref{res_EH} the shear viscosity $\eta$ for our highest temperatures.
The TTS scaling of $\GF(\tsamp,T)$ will be presented in Sec.~\ref{res_GF_TTS}.
We confirm in Sec.~\ref{res_Gt} the TTS scaling for the directly determined shear-stress relaxation 
modulus $G(t)$. That $\GF(\tsamp)$ is statistically better behaved as $G(t)$ is demonstrated 
using the standard deviations $\delta \GF$ and $\delta G$ discussed in Sec.~\ref{res_dGdmu}.
We conclude the paper in Sec.~\ref{sec_conc}.
The definitions of $\tauhat$ and $\muAhat$ are given in Appendix~\ref{app_useful}.
The model Hamiltonian is described in Appendix~\ref{app_Hamiltonian}.
Details concerning the time and ensemble averages used can be found in Appendix~\ref{app_data}.
The difference of simple averages and fluctuations is stressed in Appendix~\ref{app_simplefluctu}.
Appendix~\ref{app_fluctu} reminds briefly the derivation of the stationarity relation, 
Eq.~(\ref{eq_muGt}), already presented elsewhere \cite{WXB15,WKC16,ivan18a}.

\section{Algorithm and ensembles}
\label{sec_algo}

\begin{table}[t]
\begin{center}
\begin{tabular}{|c|c|c||c|c|c|c|c|c|c|c|}
\hline
\hline
ensemble &$L$ &$m$&$\Tglass$& $H$ 
                         &$\muA$
                               &$\muF$
                                     &$\GF$ 
                                            &$\Rgyr$
                                                 &$\Rend$
                                                       & $H/\Rgyr$ \\ \hline \hline
3D bulk& -   & 10& 0.395 & -    & 93.3& 84.6&8.7   & 1.9 & 4.6 & -    \\
\hline
film 1 & 23.5&120& 0.371 & 21.3 & 93.9& 85.6&8.3   & 1.9 & 4.6 & 11.3 \\
film 2 & 37.1& 10& 0.334 & 8.5  & 94.2& 86.1&8.1   & 1.9 & 4.6 & 4.5  \\
film 3 & 42  & 10& 0.318 & 6.6  & 94.3& 86.5&7.8   & 1.9 & 4.6 & 3.5  \\
film 4 & 49  & 10& 0.290 & 4.8  & 94.9& 87.4&7.5   & 1.8 & 4.4 & 2.6  \\
\hline \hline
\end{tabular}
\vspace*{0.5cm}
\caption[]{Some properties at the glass transition for the bulk 
and for films of different lateral box sizes $L$ ensemble-averaged
over $m$ independent configurations:
glass transition temperature $\Tglass$,
film thickness $H$, 
affine shear modulus $\muA$, 
shear-stress fluctuation $\muF$, 
shear modulus $\GF$ according to Eq.~(\ref{eq_SFF}),
radius of gyration $\Rgyr$,
end-to-end distance $\Rend$ \cite{RubinsteinBook} and ratio $H/\Rgyr$.
The bulk results have been obtained at an imposed average normal pressure
$P =0$ using cubic periodic boxes.
As emphasized in Sec.~\ref{res_tsamp}, it is important to specify that $\muF$ and $\GF$
have been obtained for a sampling time $\tsamp=10^4$.
\label{tab_Tg}}
\end{center}
\end{table}

\subsection{General simulation aspects}
\label{algo_simu}
As in our earlier work \cite{SBM11,ivan17c,ivan18a} our results are obtained by means of 
MD simulation of a coarse-grained bead-spring model of Kremer-Grest type \cite{LAMMPS}.
Details concerning the model Hamiltonian may be found in Appendix~\ref{app_Hamiltonian}.
Albeit the crossing of chains is effectively impossible in this model, entanglement effects 
are irrelevant for our short monodisperse chains of length $N=16$ considered and 
Rouse-type dynamics \cite{RubinsteinBook} is observed at high temperatures.
We use a velocity-Verlet scheme \cite{AllenTildesleyBook} with time steps of length $\dtMD = 0.005$. 
Temperature is imposed by means of the Nos\'e-Hoover algorithm provided by LAMMPS \cite{LAMMPS}.
Periodic boundary conditions \cite{AllenTildesleyBook} are used for all our ensembles.

\subsection{Film ensembles}
\label{algo_film}
We study free-standing polymer films containing $M=768$ chains.
As sketched in Fig.~\ref{fig_sketchSetup}, 
the films are suspended parallel to the $(x,y)$-plane with the same lateral box size
$L$ in $x$ and $y$ directions. As may be seen from Table~\ref{tab_Tg}, we simulate
ensembles with either 
$L=23.5$ (called ``film 1"), $L=37.1$ (``film 2"), $L=42$ (``film 3") or $L=49$ (``film 4").
The smallest $L$ corresponds to our thickest films on which the discussion will often focus. 
Ensemble averages over $m=120$ independently quenched configurations are performed for film 1, 
much more than the $m=10$ configurations considered for all other ensembles.
The vertical box size $\Lz$ is chosen sufficiently large ($\Lz \gg H$) to avoid any interaction 
in this direction. The instantaneous stress tensor \cite{AllenTildesleyBook} vanishes outside the films.
While this implies for all $z$-planes within the films that the average vertical normal stress
must vanish \cite{Varnik00}, some of the tangential normal stresses must be finite.
The surface tension $\Gamma$ \cite{AllenTildesleyBook,Varnik00} would otherwise vanish 
and the film be unstable. 
Note that $\Gamma \approx 1.7$ at the glass transition for all systems studied.
It decreases weakly with temperature, but remains of order unity for all films study. 
As clarified in Appendix~\ref{app_tauAmuA}, it is thus generally not appropriate to 
neglect the surface tension contribution to the Born-Lam\'e coefficients of stable films 
\cite{Riggleman13}.

\subsection{Bulk ensembles}
\label{algo_bulk}
For comparison we simulate in addition 3D bulk ensembles of same chain length $N$ and 
chain number $M$ contained in cubic periodic boxes at an imposed average pressure $P=0$. 
While the trace of the stress tensor must thus vanish on average for each configuration of the ensemble, 
this does not mean that the vertical normal stress for each $z$-pane must vanish. 
This only applies for the ensemble average over independent configurations.
This may matter (at least in principle) below the glass transition where frozen 
out-of-equilibrium stresses appear \cite{ivan18a}.
The ensembles used for bulk and film systems are thus similar, but not exactly identical 
as it would have been the case by imposing a vanishing normal stress in the $z$-direction 
at a constant linear box length $L$ in $x$- and $y$-directions. As shown in Sec.~\ref{sec_res},
this difference appears, however, not to matter: all film data extrapolates nicely to the bulk data 
(indicated by stars) if plotted as a function of $1/H$ and assuming that the bulk data corresponds 
formally to the limit $1/H=0$.

\subsection{Quench protocol and data sampling}
\label{algo_quench}
As already pointed out (Sec.~\ref{algo_film})
we do not directly vary the film thickness $H$, 
but rather impose the lateral box width $L$.
We first equilibrate an ensemble of $m$ independent films at $T=0.7$.
As shown in Sec.~\ref{res_density}, this temperature is well above the glass transition temperature 
$\Tglass$ of all systems. We then quench each configuration using a constant quench rate.
Specifically, we impose $T(t) = 0.7 - 2 \cdot 10^{-5} t$.
Fixing then a constant temperature each configuration is first tempered over a time interval 
$\ttemp = 10^5$. The subsequent production runs are performed over $\tsampmax=10^5$.
The same quench and production protocols are used for films and 3D bulk systems.
Details concerning the different types of averages sampled can be found in Appendix~\ref{app_data}
and Appendix~\ref{app_simplefluctu}.
See Table~\ref{tab_Tg} for several properties obtained at the glass transition. 

\section{Numerical results}
\label{sec_res}
\begin{figure}[t]
\centerline{\resizebox{1.0\columnwidth}{!}{\includegraphics*{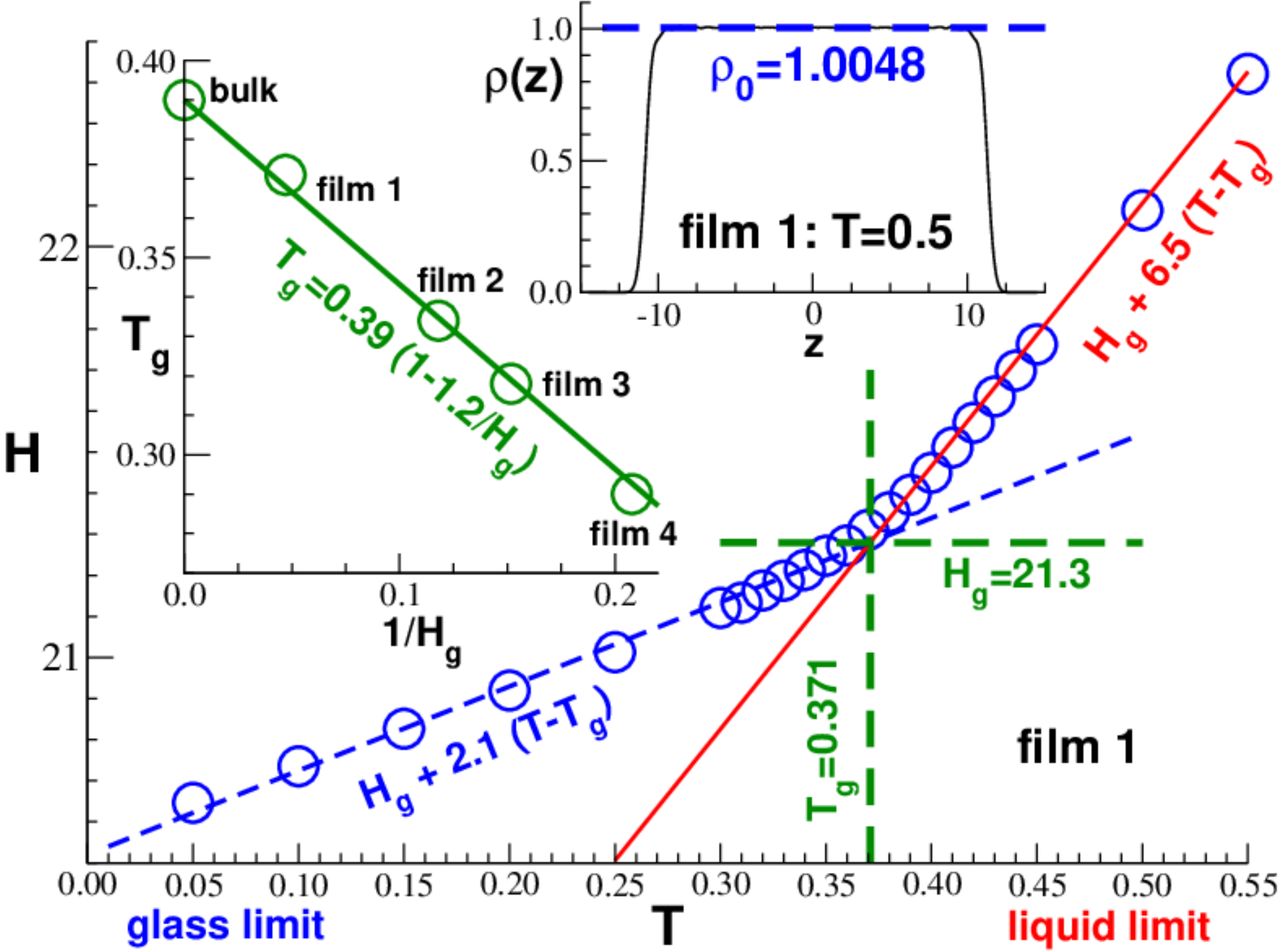}}}
\caption{Film thickness and glass transition temperature.
Top inset: Number density profile $\rho(z)$ for $T=0.5$ with $z=0$
corresponding to the center of mass of each film. 
The midplane density $\rho_0 \approx 1$ is indicated by the dashed horizontal line.
Main panel: $H$ as a function of temperature $T$ for film 1.
The glass transition temperature $\Tglass$ and the film thickness $\Hglass$
at the transition (bold dashed lines) are operationally defined by the intercept of the
linear extrapolations of the glass (dashed line) and liquid (solid line) limits.
Left inset: $\Tglass$ as a function of $1/\Hglass$ confirming
the linear superposition, Eq.~(\ref{eq_Acalsurfeff}).
}
\label{fig_H_T}
\end{figure}

\subsection{Film thickness and glass transition temperature}
\label{res_density}

%
A central parameter for the description of our films is the film thickness $H$.
We determine $H$ using a Gibbs dividing surface construction \cite{PlischkeBook,Pablo00}. 
With $\rho_0 \equiv \rho(z \approx 0)$ being the midplane density of the density profile $\rho(z)$,
this implies
\begin{equation}
H \equiv N M / \rho_0 L^2.
\label{eq_H}
\end{equation}
As seen for one example in the top inset of Fig.~\ref{fig_H_T}, 
$\rho(z)$ is always uniform and smooth around the midplane 
in agreement with the data presented in previous studies \cite{Pablo02b}. 
$\rho_0$ can thus be fitted to high precision and, hence, also $H$.
Since $\rho_0$ is always very close to unity, varying only little with $L$, 
Eq.~(\ref{eq_H}) implies that (to leading order) $H \sim 1/L^2$ changes very strongly with $L$.

We present in the main panel of Fig.~\ref{fig_H_T} the film thickness as a function of temperature. 
As emphasized by the dashed and the solid lines, the film thickness $H$ 
--- and thus the film volume $V=L^2H$ ---
decreases monotonically upon cooling with the two linear branches fitting reasonably
the glass (dashed line) and the liquid (solid line) limits. 
The intercept (horizontal and vertical dashed lines) of both asymptotes allows to define 
the apparent glass transition temperature $\Tglass$ and the film thickness $\Hglass$ 
at the transition \cite{Pablo02b}. (See Ref.~\cite{ivan18a} for bulk systems.)
The values are given in Table~\ref{tab_Tg}.

As expected from a wealth of literature  
\cite{Dalnoki96,Forrest00,Dalnoki01,Dalnoki12,Dalnoki14,Pablo00,Pablo02,Pablo02b,Varnik02,Peter06,Peter07,LangSimmons:Macro2013}, 
$\Tglass$ increases with $H$. More precisely, as seen in the left inset of Fig.~\ref{fig_H_T}, 
$\Tglass$ extrapolates linearly with the inverse film thickness to the thick-film limit.
(The value $\Tglass=0.395$ indicated at $1/\Hglass=0$ stems from our bulk simulations.) 
This is consistent with a linear superposition, Eq.~(\ref{eq_Acalsurfeff}), 
of a thickness-independent bulk glass transition temperature $T_{\rm g0}$ and 
an effective surface temperature $T_{\rm gs}$ \cite{foot_otherH}. 
The negative sign of the correction implies $T_{\rm gs} < T_{\rm g0}$,
i.e. surface relaxation processes are faster than processes around the film midplane. 
This is consistent with the higher monomer mobilities observed at the film surfaces 
\cite{KTT95,Pablo00,Pablo02,Peter06,YangEtal:Science2010,LamTsui:PRE2013,ChowdhuryEtal:JPCL2017}.
We emphasize finally that many more data points covering a much broader range of orders of magnitude 
in $1/H$ are required to find or to rule out numerically higher orders of a systematic $1/H$-expansion 
of $\Tglass$.

\begin{figure}[t]
\centerline{\resizebox{1.0\columnwidth}{!}{\includegraphics*{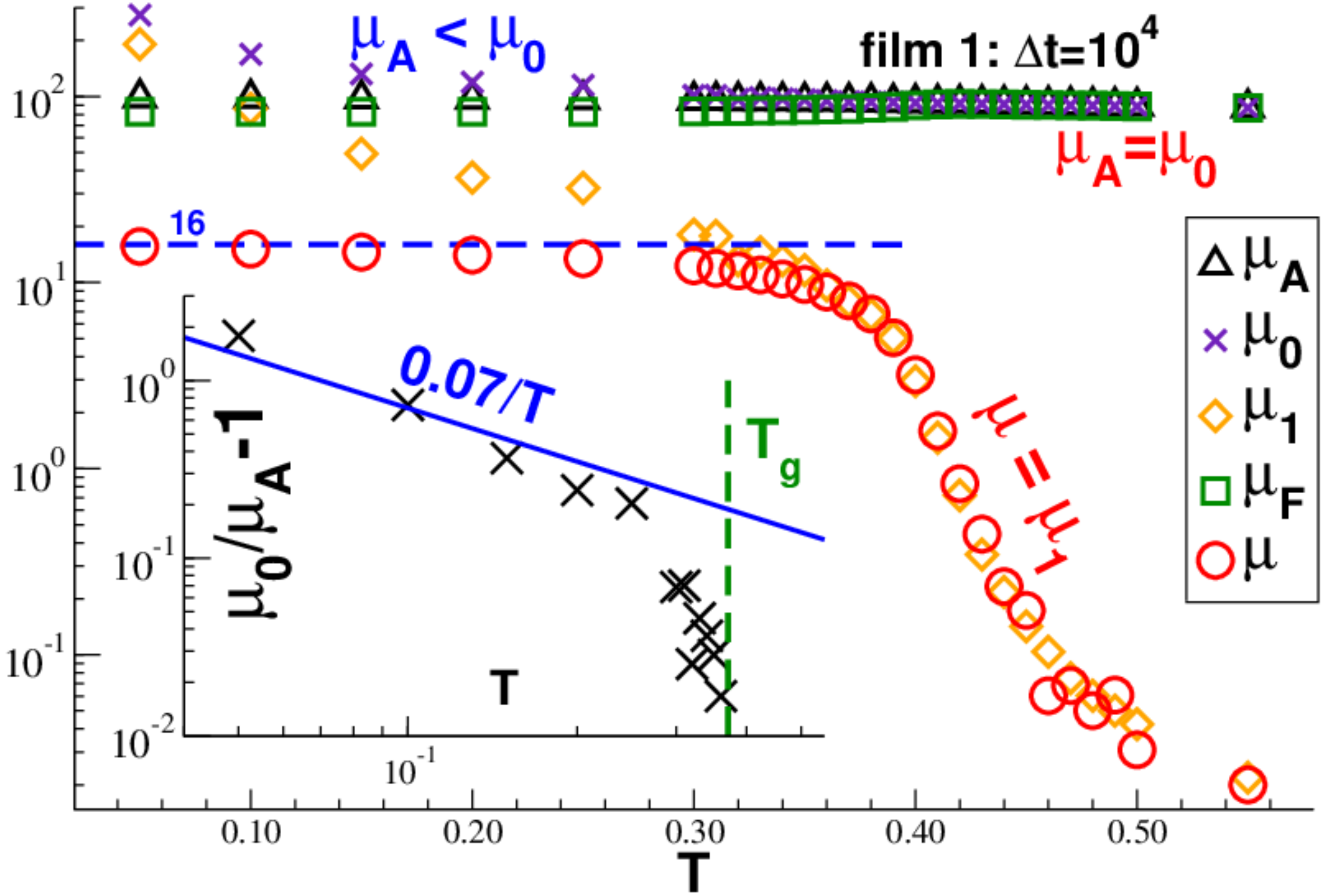}}}
\caption{Comparison of the different contributions to the shear modulus 
$\GF = \muA-\muF=(\muA-\muFtwo)+\muFone$
as functions of $T$ focusing on data obtained for film 1 and $\tsamp=10^4$.
Inset: Double-logarithmic representation of $\muFtwo/\muA-1$ {\em vs} $T$.
}
\label{fig_SFF}
\end{figure}

\subsection{Stress-fluctuation formula at constant $\tsamp$}
\label{res_SFF}

Instantaneous values of the shear stress $\tauhat$ and of the affine shear modulus $\muAhat$ 
have been computed as described in Appendix~\ref{app_tauAmuA}.
The time and ensemble averaged affine shear modulus  $\muA \equiv \langle \overline{\muAhat} \rangle$
is presented in Fig.~\ref{fig_SFF} as a function of temperature using half-logarithmic coordinates.
The averaged shear stress $\sxy \equiv \langle \overline{\tauhat} \rangle$ is not indicated
since it vanishes rapidly due to symmetry with increasing ensemble size $m$ and sampling time $\tsamp$.
As seen from Fig.~\ref{fig_SFF}, this is not the case for the moments
\begin{equation}
\muFtwo \equiv \beta V \la \overline{ \tauhat^2 } \ra,
\muFone \equiv \beta V \la \overline{\tauhat}^2 \ra,
\muF \equiv \muFtwo - \muFone
\label{eq_taumoments}
\end{equation}
(with $\beta=1/T$ being the inverse temperature) 
describing the non-affine contributions to the stress-fluctuation formula Eq.~(\ref{eq_SFF}).
Note that $\muF$, $\muFtwo$ and $\muA$ depend only weakly on $T$ and are all similar
on the logarithmic scale used in Fig.~\ref{fig_SFF}.
 
%
As stressed elsewhere \cite{ivan18a}, $\muA=\muFtwo$ for an equilibrium liquid 
since both $\GF=(\muA-\muFtwo)+\muFone$ and $\muFone$ must vanish. 
Frozen-in out-of-equilibrium stresses are observed upon cooling below $\Tglass$
as made manifest by the dramatic increase of the dimensionless ratio $\muFtwo/\muA-1$.
The $\beta$-prefactor of $\muFtwo$, Eq.~(\ref{eq_taumoments}), implies that due to the frozen stresses 
\begin{equation}
\muFtwo/\muA-1 \sim 1/T \mbox{ for } T \ll \Tglass
\label{eq_muFtwomuA}
\end{equation}
to leading order.
This is consistent with the data presented in the inset of Fig.~\ref{fig_SFF}.
Similar behavior has been reported for 3D polymer bulks \cite{ivan18a}.

\begin{figure}[t]
\centerline{\resizebox{1.0\columnwidth}{!}{\includegraphics*{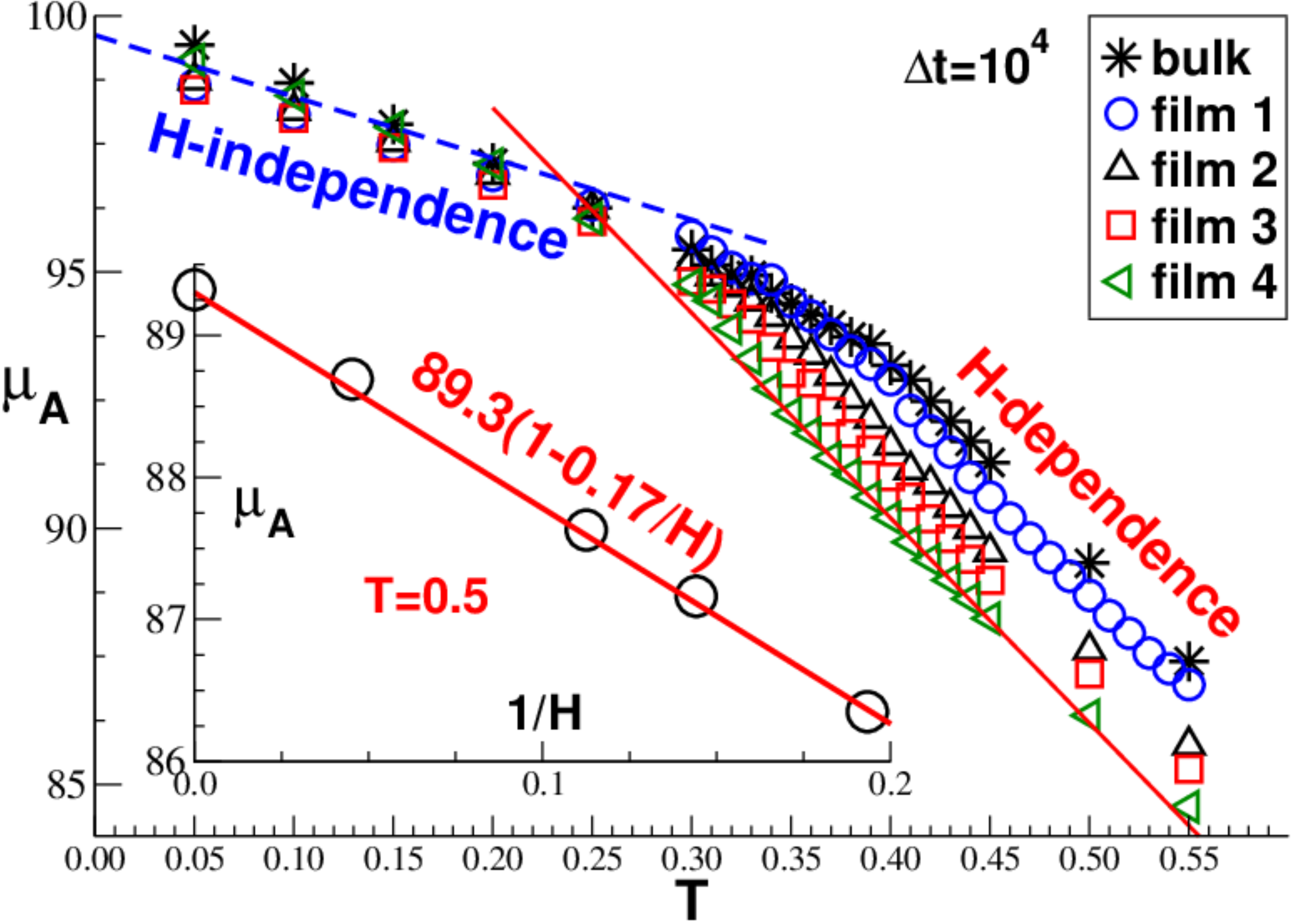}}}
\caption{Affine shear modulus $\muA$.
Main panel: $\muA(T)$ for all systems studied.
Inset: As shown for $T=0.5$, $\muA$ decreases linearly with $1/H$ in the liquid limit.
}
\label{fig_muA_T}
\end{figure}

%
Using a linear representation, the main panel of Fig.~\ref{fig_muA_T} presents $\muA(T)$ for all ensembles.
This shows (more clearly than Fig.~\ref{fig_SFF}) that
$\muA$ decreases continuously with temperature with two (approximately) linear
branches in the glass and the liquid regimes as indicate by the two lines.
While $\muA$ barely depends on $H$ in the glass limit
(suggesting a weak surface contribution $\muAsurf$),
it increases with $H$ in the liquid limit. 
As demonstrated in the inset, $\muA$ decreases in fact linearly with $1/H$ 
in agreement with Eq.~(\ref{eq_Acalsurfeff}) \cite{foot_muAz}.

\begin{figure}[t]
\centerline{\resizebox{1.0\columnwidth}{!}{\includegraphics*{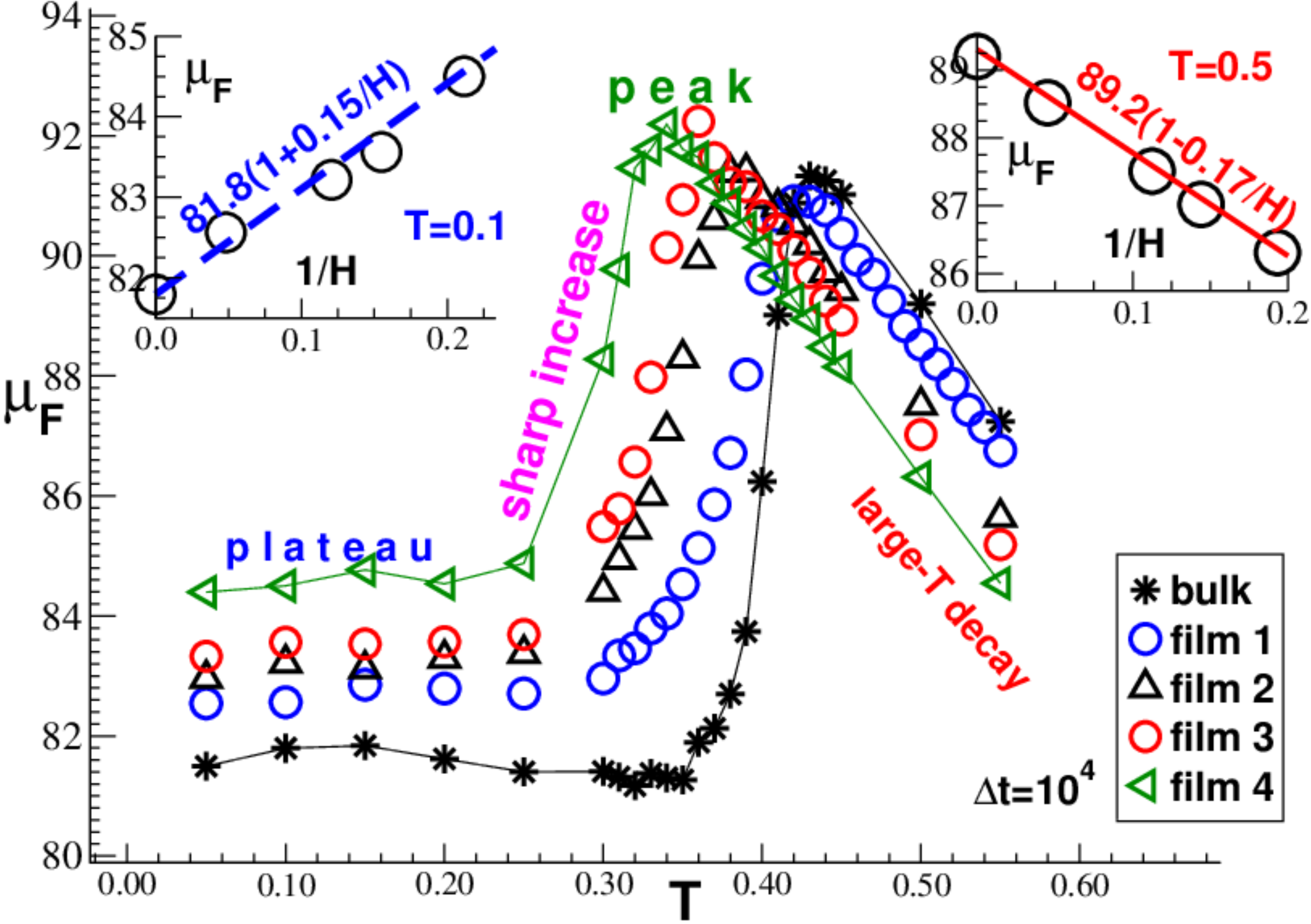}}}
\caption{Shear-stress fluctuation $\muF$ for $\tsamp=10^4$.
Main panel: $\muF(T)$ for all systems.
Right inset:
$\muF$ decreases linearly with $1/H$ in the liquid limit ($T=0.5$).
Left inset:
$\muF$ increases linearly with $1/H$ in the solid limit ($T=0.1$).
}
\label{fig_muF_T}
\end{figure}

%
Using again a linear representation $\muF(T)$ is presented in the main panel of Fig.~\ref{fig_muF_T}. 
Upon cooling it increases first (essentially linearly), goes through a 
well-defined peak located around $\Tglass$ and drops then rapidly albeit continuously.
It becomes constant for $T \ll \Tglass$ when the shear stresses get quenched.
Since $\muA \approx \muF$ at high temperatures, the same linear $1/H$-dependences
are naturally observed as shown in the right inset of Fig.~\ref{fig_muF_T} for $T=0.5$.
At variance to this, $\muF$ {\em increases} linearly with $1/H$ at low temperatures
as seen for $T=0.1$ in the left inset, i.e. the non-affine contributions are the
largest for our thinnest films.
Both linear $1/H$-relations for $\muF$ are consistent with Eq.~(\ref{eq_Acalsurfeff}). 
The negative sign of the correction for large $T$ suggests that the bulk value $\muFbulk$ in the middle 
of the films must exceed the value $\muFsurf$ at the surfaces while the opposite behavior
occurs in the low-$T$ limit \cite{foot_muFsign}.

\begin{figure}[t]
\centerline{\resizebox{1.0\columnwidth}{!}{\includegraphics*{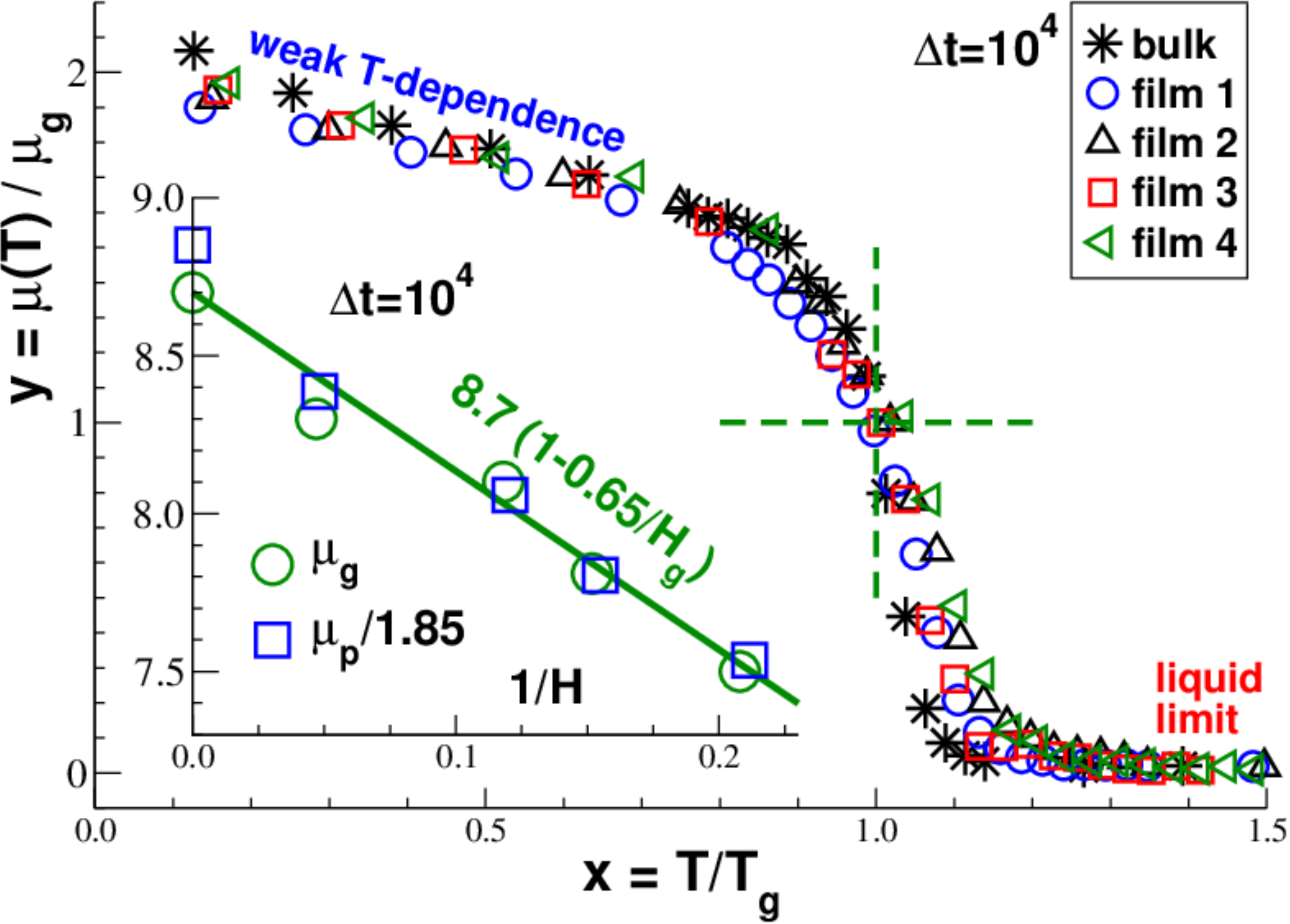}}}
\caption{Film thickness dependence of $\GF(T)$ for $\tsamp=10^4$.
Main panel: Scaling collapse of $y=\GF(T)/\GFglass$ {\em vs} $x=T/\Tglass$.
Inset: $\GFglass \equiv \GF(\Tglass)$ and $\GFplat\equiv \GF(T=0.1)$ {\em vs} 
the inverse film thickness $1/H$ of the respective temperature.
As emphasized by the bold line, both shear moduli are consistent with Eq.~(\ref{eq_Acalsurfeff}). 
We shall use $\GFplat$ in Sec.~\ref{res_GF_TTS} and Sec.~\ref{res_Gt} for the TTS scaling of 
$\GF(\tsamp)$ and $G(t)$ comparing different ensembles. 
}
\label{fig_GF_Lscal}
\end{figure}

%
As already highlighted in the main panel of Fig.~\ref{fig_key}, the shear modulus $\GF$ depends on 
the film thickness just as its affine (Fig.~\ref{fig_muA_T}) and  non-affine (Fig.~\ref{fig_muF_T})
contributions. 
Focusing on $\GF(T)$ it is shown in the main panel of Fig.~\ref{fig_GF_Lscal} 
that these properties can be brought to collapse on $H$-independent mastercurves.
The horizontal axis is rescaled with the reduced temperature $T/\Tglass$ using the apparent 
glass transition temperature $\Tglass$ defined in Sec.~\ref{res_density}.
The values $\GFglass \equiv  \GF(\Tglass)$ used to make the vertical axes dimensionless 
are indicated in Table~\ref{tab_Tg} and plotted in the inset of Fig.~\ref{fig_GF_Lscal}.
Consistently with the linear superposition relation, 
$\GFglass$ is a linear function of $1/\Hglass$. 
Similar scaling plots could be given for the contributions
$\muA(T)$, $\muFtwo(T)$, $\muFone(T)$ and $\muF(T)$.

\begin{figure}[t]
\centerline{\resizebox{1.0\columnwidth}{!}{\includegraphics*{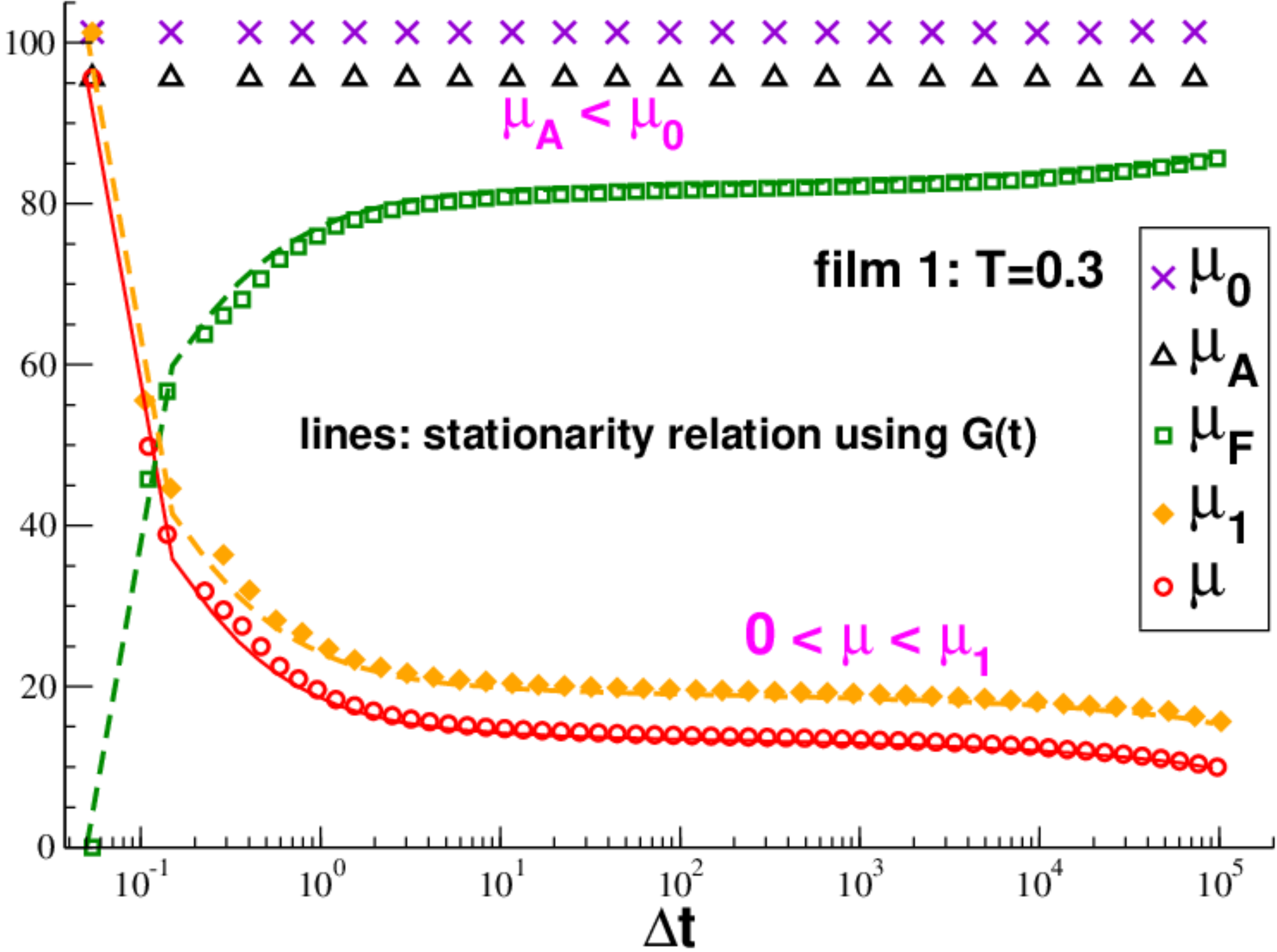}}}
\caption{Sampling time effects for $\GF$ and its contributions focusing on film 1 and $T=0.3$.
Only the simple averages $\muA$ and $\muFtwo$ are strictly $\tsamp$-independent.
$\muFone$ and (hence) $\GF$ decrease monotonically.
The solid and dashes lines have been obtained using Eq.~(\ref{eq_muGt}).
}
\label{fig_tsamp}
\end{figure}

\subsection{Effective time-translational invariance}
\label{res_tsamp}

All data presented in the previous subsection have been obtained for one sampling time $\tsamp=10^4$.
We turn now to the characterization of the $\tsamp$-effects observed for $\GF$ 
in the inset of Fig.~\ref{fig_key}. Focusing on one temperature ($T=0.3$) in the 
glass limit, we compare in Fig.~\ref{fig_tsamp} the $\tsamp$-dependencies of
$\muA$, $\muFtwo$, $\muFone$, $\muF$ and $\GF$. 
As expected from Eq.~(\ref{eq_commute}), the simple averages $\muA$ and $\muFtwo$
are found to be strictly $\tsamp$-independent. 
Importantly, time and ensemble averages do not commute for $\muFone$ since
\begin{equation}
0 = \beta V \overline{ \langle \tauhat \rangle^2} < 
\beta V \langle \overline{\tauhat}^2 \rangle \equiv \muFone(\tsamp),
\end{equation}
i.e. $\muFone$ is not a simple average, but a fluctuation.
As seen in Fig.~\ref{fig_tsamp}, $\muFone(\tsamp)$ decays in fact monotonically
and, as a consequence, $\muF(\tsamp) = \muFtwo-\muFone(\tsamp)$ increases
and $\GF(\tsamp)=(\muA-\muFtwo)+\muFone(\tsamp)$ decreases monotonically.
Interestingly, as indicated by the thin solid line, the stationarity relation Eq.~(\ref{eq_muGt}) holds,
i.e. $\GF(\tsamp)$ can be traced back from the independently determined shear-stress relaxation modulus 
$G(t)$ discussed in Sec.~\ref{res_Gt}.
(The visible minor differences are due to numerical difficulties related to the finite time step 
and the inaccurate integration of the strongly oscillatory $G(t)$ at short times.)
Since $\muA$ and $\muFtwo$ are $\tsamp$-independent simple averages,
one can rewrite Eq.~(\ref{eq_muGt}) to also describe $\muFone(\tsamp)$ and $\muF(\tsamp)$.
This is indicated by the two dashed lines. Note that Eq.~(\ref{eq_muGt}) 
has been shown to hold for all temperatures and ensembles.
The observed $\tsamp$-dependence of the shear modulus $\GF$ is thus not due to aging effects, 
but arises naturally from the effective time translational invariance of our systems.
This does, of course, not mean that no aging occurs in our glassy systems,
but just that this is irrelevant for the time scales and the properties considered here.
We shall now use the decay of $\GF(\tsamp) \approx \muFone(\tsamp)$ for large $T$ and $\tsamp$
to characterize the shear viscosity $\eta(T)$. 

\subsection{Plateau modulus $\GFplat$ and shear viscosity $\eta$}
\label{res_EH}

\begin{figure}[t]
\centerline{\resizebox{1.0\columnwidth}{!}{\includegraphics*{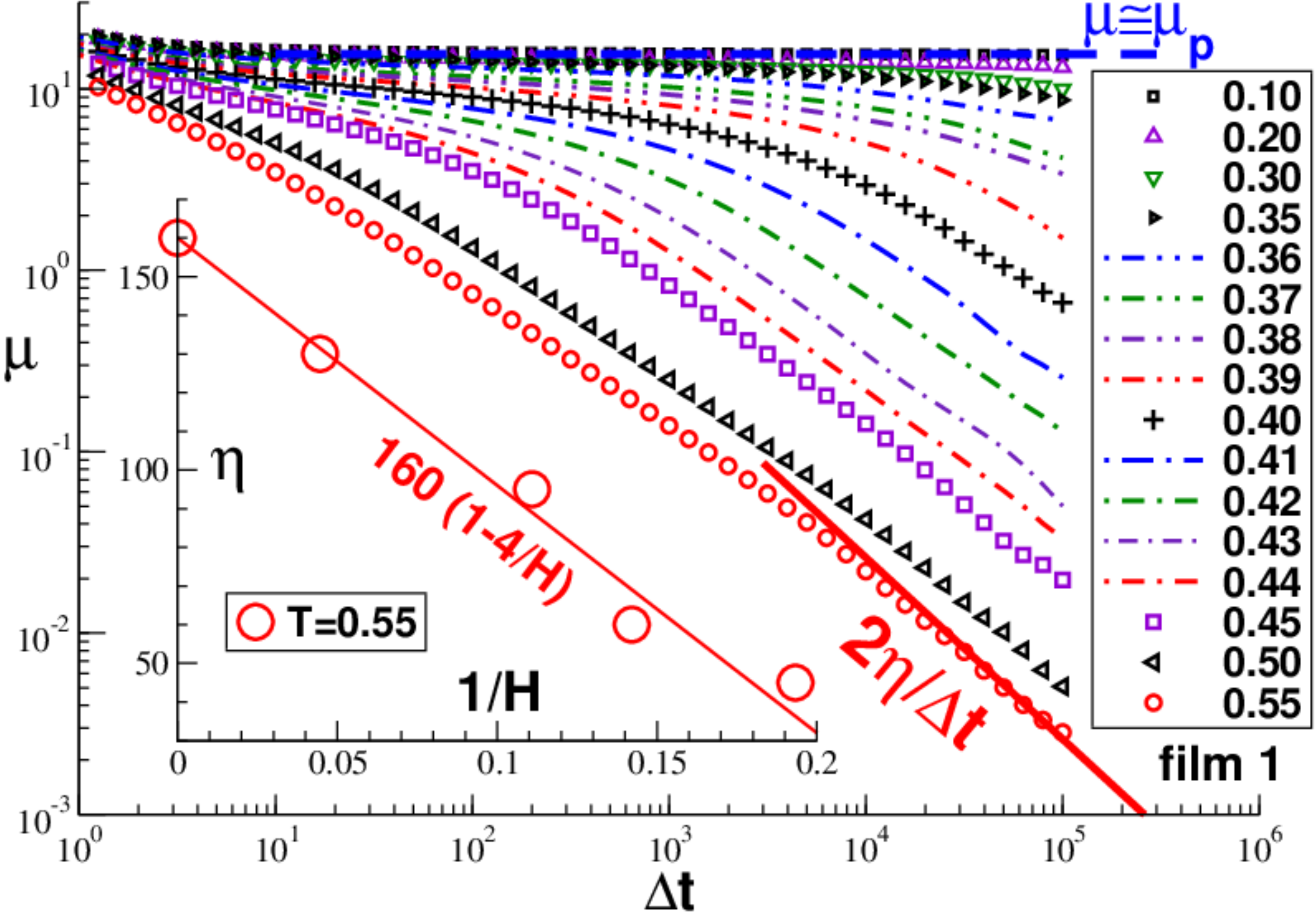}}}
\caption{Double-logarithmic representation of $\GF(\tsamp)$ for 
a broad range of temperatures $T$ focusing on film 1. 
$\GF(\tsamp;T)$ decreases continuously with both $\tsamp$ and $T$.
A pseudo-elastic plateau is observed in the solid limit with 
$\GF \approx \GFplat \approx 15.5$ (horizontal dashed line).
The $1/\tsamp$-decay in the liquid limit (bold solid line) is expected from 
the Einstein-Helfand relation, Eq.~(\ref{eq_EH}). 
Inset: Shear viscosity $\eta(1/H)$ for $T=0.55$.
The values are used in Sec.~\ref{res_GF_TTS} to define an absolute scale for $\tauterm(T)$.
The line presents a linear fit according to Eq.~(\ref{eq_Acalsurfeff}).
}
\label{fig_EH}
\end{figure}

That $\GF$ decreases monotonically with $\tsamp$ is also seen in the 
main panel of Fig.~\ref{fig_EH} for a broad range of temperatures using a double-logarithmic representation.
As already pointed out above (Fig.~\ref{fig_key}), it also decreases continuously with $T$ and no indication 
of a jump singularity is observed.
We emphasize that the same qualitative behavior is found for all systems we have investigated.
(Similar plots have been obtained for glass-forming colloids in 2D \cite{WXP13} and for 
3D polymers \cite{ivan17c,ivan18a}.)

As one expects, the $\tsamp$-dependence of $\GF$ becomes extremely weak in the solid limit,
i.e. a plateau (shoulder) $\GF(\tsamp) \approx \GFplat=const$ appears for a broad $\tsamp$-window.
Since the plateau value $\GFplat$ depends somewhat on $T$ and on the $\tsamp$-window fitted,
it is convenient for the dimensionless scaling plots presented in the next two 
subsections to define $\GFplat(H) \equiv \GF(T=0.1,\tsamp=10^4,H)$.
The value for film 1 is indicated by the horizontal dashed line.
As may be seen from the inset of Fig.~\ref{fig_GF_Lscal}, 
\begin{equation}
\GFplat(H) \approx 16.1 \ (1-0.65/H) \approx 1.85 \GFglass(H)
\label{eq_GFplat}
\end{equation}
in agreement with Eq.~(\ref{eq_Acalsurfeff}).

As emphasized by the bold solid line in the main panel of Fig.~\ref{fig_EH}, 
$\GF(\tsamp)$ decreases inversely with $\tsamp$ in the high-$T$ limit. 
This is expected from the Einstein-Helfand relation \cite{AllenTildesleyBook,ivan18a}
\begin{equation}
\GF(\tsamp) \to 2\eta/\tsamp \mbox{ for } \tsamp \gg \tauterm
\label{eq_EH}
\end{equation}
with $\eta$ being the shear viscosity and $\tauterm$ the terminal shear stress relaxation time.
Note that Eq.~(\ref{eq_EH}) follows directly from the stationarity relation 
Eq.~(\ref{eq_muGt}) and the more familiar Green-Kubo relation 
$\eta = \int_0^{\infty} \ddiff t  \ G(t)$ 
for the shear viscosity \cite{RubinsteinBook}.
A technical point must be mentioned here. 
We remind that $\muA=\muFtwo$ in the liquid limit implies $\GF(\tsamp)=\muFone(\tsamp)$. 
Since the impulsive corrections needed for the calculation of $\muA$ and, hence, of $\GF$ 
are not sufficiently precise for the logarithmic scale used here, it is for numerically 
reasons best to simply replace $\GF$ by $\muFone$ to avoid an artificial curvature of the
data for large $\tsamp$. (See Fig.~16 of Ref.~\cite{ivan18a} for an illustration.)
Using the Einstein-Helfand relation it is then possible to fit $\eta(T)$ above $T \approx 0.5$. 
For smaller temperatures this method only allows the estimation of lower bounds. 
(See the inset (b) of Fig.~17 of Ref.~\cite{ivan18a} for 3D bulks.)
As shown in the inset of Fig.~\ref{fig_EH} for $T=0.55$, the shear viscosity decreases
systematically for thinner films and the linear superposition relation (solid line) 
describes reasonably all available data.
We show now how $\eta(T)$ may be extrapolated to much smaller temperatures by means of the 
TTS scaling of $\GF(\tsamp)$.


\begin{figure}[t]
\centerline{\resizebox{1.0\columnwidth}{!}{\includegraphics*{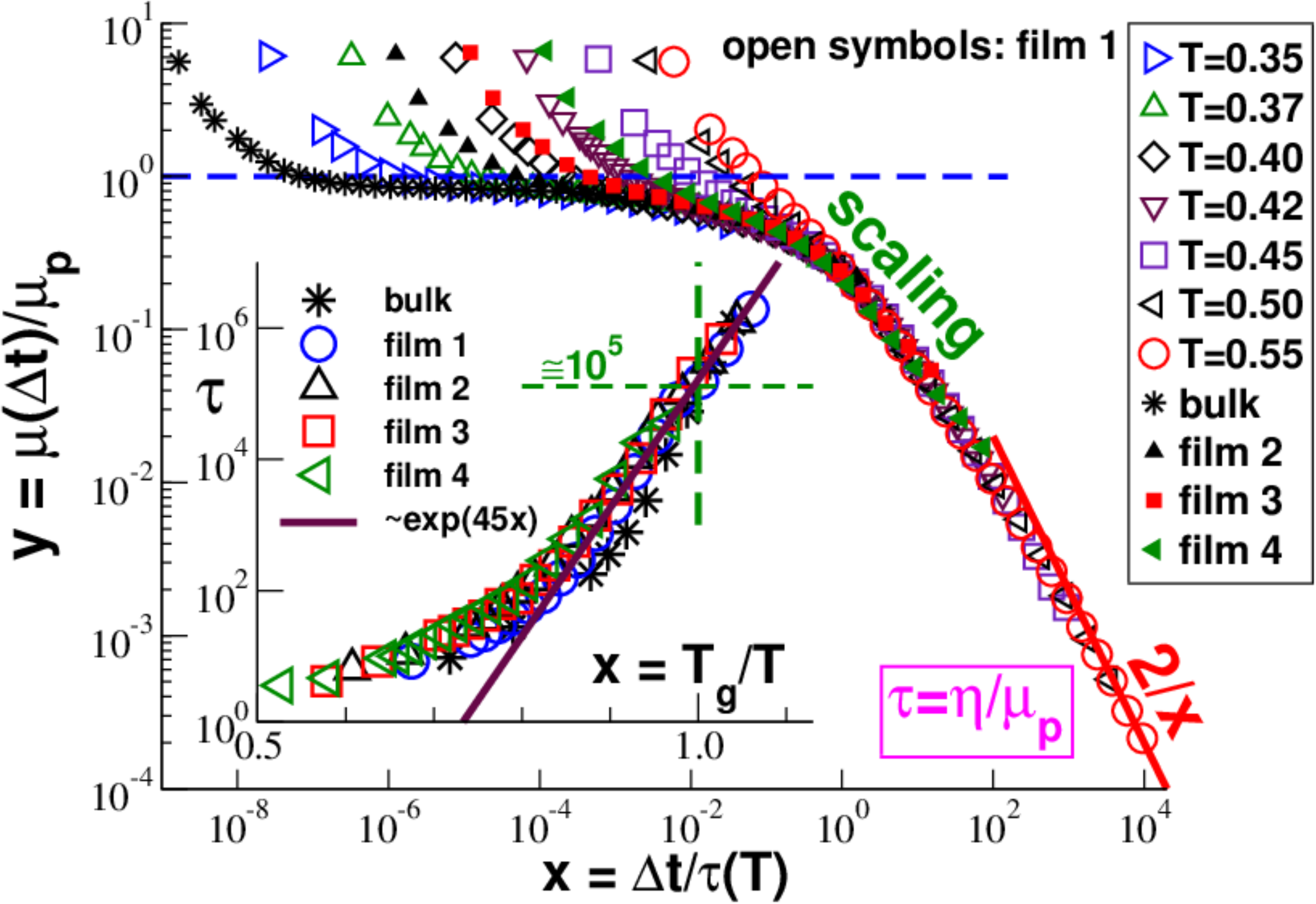}}}
\caption{TTS scaling for $y = \GF(\tsamp)/\GFplat$ as a function of $x = \tsamp/\tauterm(T)$
with $\GFplat$ being the plateau modulus defined in Sec.~\ref{res_EH} and 
$\tauterm(T)$ the relaxation time indicated in the inset. 
We impose $\tauterm(T=0.55)$ according to Eq.~(\ref{eq_taueta}) to have an absolute time scale.
The two asymptotics of the scaling function $y=f(x)$ for $x \ll 1$ and
$x \gg 1$ are indicated by dashed and solid lines.
Note the broad crossover regime between these limits.
Inset: 
Data collapse of terminal relaxation time $\tauterm$ {\em vs} $x=\Tglass/T$ for all our ensembles.
Arrhenius behavior (bold solid line) is observed around the glass transition ($x \approx 1$).
}
\label{fig_GF_TTS}
\end{figure}

\subsection{Time-temperature superposition of $\GF(\tsamp)$}
\label{res_GF_TTS}
The TTS scaling of $\GF(\tsamp)$ is presented in the main panel of Fig.~\ref{fig_GF_TTS}
using dimensionless coordinates and a double-logarithmic representation.
Data for a broad range of temperatures are given for film 1 (open symbols) while we focus 
for clarity on one temperature ($T=0.35$) for the other films (filled symbols) and 
the 3D bulk ensembles (stars). A good data collapse is achieved by plotting the rescaled 
shear modulus $y = \GF(\tsamp)/\GFplat$ as a function of the reduced sampling time 
$x = \tsamp/\tauterm(T)$ using the relaxation time $\tauterm(T)$ indicated in the inset. 
The scaling function $y=f(x)$ is given by $f(x) \to const \approx 1$ for $x \ll 1$ (dashed horizontal line)
and by $f(x) \to 2/x$ for $x \gg 1$ (bold solid line) for consistency with the Einstein-Helfand relation. 
The vertical axis is made dimensionless using the plateau modulus $\GFplat$ introduced in Sec.~\ref{res_EH}. 
Please note that since according to Eq.~(\ref{eq_GFplat}) the $H$-dependence of $\GFplat$ 
is rather small on the logarithmic scales we are interested in, a similar good data collapse 
may also be achieved by simply setting $\GFplat=1$.
Much more important is the rescaling of the horizontal axis by means of the terminal relaxation 
time $\tauterm(T,H)$ which depends strongly on both temperature and film thickness.
Note that the strong $H$-dependence is masked by the rescaling of the horizontal axis
using $x=\Tglass(H)/T$ in the inset of Fig.~\ref{fig_GF_TTS}.

Some remarks may be in order to explain how the scaling plot was achieved.
We have in fact followed in a first step the standard prescription \cite{FerryBook,RubinsteinBook} 
fitting the relative dimensionless factors $a_T$ and $b_T$ for the horizontal and vertical rescaling of 
$\GF(\tsamp,T)$ for temperatures $T$ close to certain reference temperatures $T_0$. 
As one may expect \cite{FerryBook}, $b_T$ can safely be set to unity for the entire 
temperature range we are interested in. In turn this justifies the temperature independent 
factor $\GFplat$ used to rescale the vertical axis.
Naturally, merely tuning $a_T=\tauterm(T)/\tauterm(T_0)$ only sets the relative 
scale of $\tauterm(T)$. In order to fix the missing prefactor we impose 
\begin{equation}
\tauterm(T) = c \ \eta(T)/\GFplat(H) \mbox{ with } c = 1 
\label{eq_taueta}
\end{equation}
for $T=T_0=0.55$
using the shear viscosity $\eta$ determined in the high-$T$ limit by means of Eq.~(\ref{eq_EH}).
Due to the somewhat arbitrary constant $c/\GFplat$ the strongest curvature of the rescaled 
shear modulus $y(x)$ coincides with $x \approx 1$.
(Using instead $c \approx 100$ the crossover to the Einstein-Helfand decay 
would occur at about $x \approx 1$.)
Consistency of $\GF(\tsamp) = \GFplat f(x) \approx \GFplat \tauterm/\tsamp$ for $x \gg 1$ 
and the Einstein-Helfand relation, 
Eq.~(\ref{eq_EH}), implies interestingly that Eq.~(\ref{eq_taueta}) must hold for all 
temperatures. In other words, the relaxation time $\tauterm(T)$, shown in the inset of 
Fig.~\ref{fig_GF_TTS}, and the shear viscosity $\eta(T)$ are equivalent up to a 
trivial prefactor.
We emphasize that the stated proportionality hinges on the observation that $b_T \approx 1$.

As shown in the inset, a remarkable scaling collapse is achieved by plotting $\tauterm$ or $\eta$ 
as a function of $x=\Tglass/T$. Especially, this implies that we find 
\begin{equation}
\tauterm(T \approx \Tglass) = c \ \eta(T \approx \Tglass)/\GFplat(H) \approx 10^5
\label{eq_tauTg}
\end{equation}
for all our ensembles as shown by the horizontal and vertical dashed lines.
In other words, the dilatometric criterion (Sec.~\ref{res_density}) and the 
rheological criterion, fixing a characteristic viscosity for defining $\Tglass$ 
\cite{FerryBook}, are numerically consistent on the logarithmic scales considered here. 
Anticipating better statistics and longer production runs (improving thus the precision of the TTS scaling),
this suggests that Eq.~(\ref{eq_tauTg}) may be used in the future to define $\Tglass$.
We note finally that an Arrhenius behavior $\tauterm \sim \exp(45x)$ is observed 
for $x \approx 1$ (bold solid line) and that higher temperatures are consistent 
with a Vogel-Fulcher-Tammann law \cite{FerryBook} (not shown).

\subsection{Shear-stress relaxation modulus $G(t)$}
\label{res_Gt}
While the (shear strain) creep compliance $J(t)$ \cite{FerryBook} of polymer films has been
obtained experimentally (by means of a biaxial strain experiment using effectively the 
reasonable approximation of a time-independent Poisson ratio near $1/2$) 
\cite{McKenna05,McKenna08,McKenna17,BodiguelFretigny:EPJE2006},
this seems not to be the case for the shear-stress relaxation modulus $G(t)$.
This could in principle be done by suddenly tilting the frame on which a free-standing film 
is suspended and by measuring the shear stress $\sigma(t)$ needed to keep constant the tilt angle $\gamma$
as shown in Fig.~\ref{fig_sketchSFF}.
The direct numerical computation of $G(t)$ by means of an out-of-equilibrium simulation 
tilting the simulation box in a similar manner, is a feasible procedure in principle as 
shown in Ref.~\cite{Pablo05b}. 
For general technical reasons \cite{AllenTildesleyBook} this procedure remains tedious, however.
(Being currently still limited to the high-frequency limit, it is especially not 
possible to get $G(t)$ by Fourier transformation of the storage and loss moduli $\Gstor$ and $\Gloss$ 
obtained by applying an oscillatory simple shear \cite{Pablo05b}.)
Fortunately, $G(t)$ can be computed ``on the fly" using the stored time-series
of $\tauhat$ and $\muAhat$ by means of the appropriate
linear-response fluctuation-dissipation relation.
It is widely assumed \cite{AllenTildesleyBook} that $G(t)$ is given  
by the shear-stress autocorrelation function 
\begin{equation}
c(t) \equiv \beta V \la \overline{\tauhat(t) \tauhat(0)} \ra.
\label{eq_ct}
\end{equation}
However, as emphasized elsewhere \cite{ivan18a}, this expression can only be used 
under the condition that $\muA=\muFtwo$. Albeit this does hold in the liquid
limit of our films, as we have seen above (Fig.~\ref{fig_SFF}), this 
condition may not be satisfied below $\Tglass$ \cite{ivan18a}. 
It is thus necessary to obtain $G(t)$ below $\Tglass$
using more generally \cite{ivan17c,ivan18a}
\begin{eqnarray}
G(t) & = & \muA - h(t) = (\muA-\muFtwo) + c(t) \mbox{ with } \label{eq_Gtgeneral} \\
h(t) & \equiv & \frac{\beta V}{2}  \ \la \overline{(\tauhat(t)-\tauhat(0))^2} \ra
= c(0)-c(t)
\label{eq_ht}
\end{eqnarray}
being the shear-stress mean-square displacement.
Note that $G(t=0)=\muA$ as it should if an affine strain
is applied at $t=0$ as sketched in panel (b) of Fig.~\ref{fig_sketchSFF}.

\begin{figure}[t]
\centerline{\resizebox{1.0\columnwidth}{!}{\includegraphics*{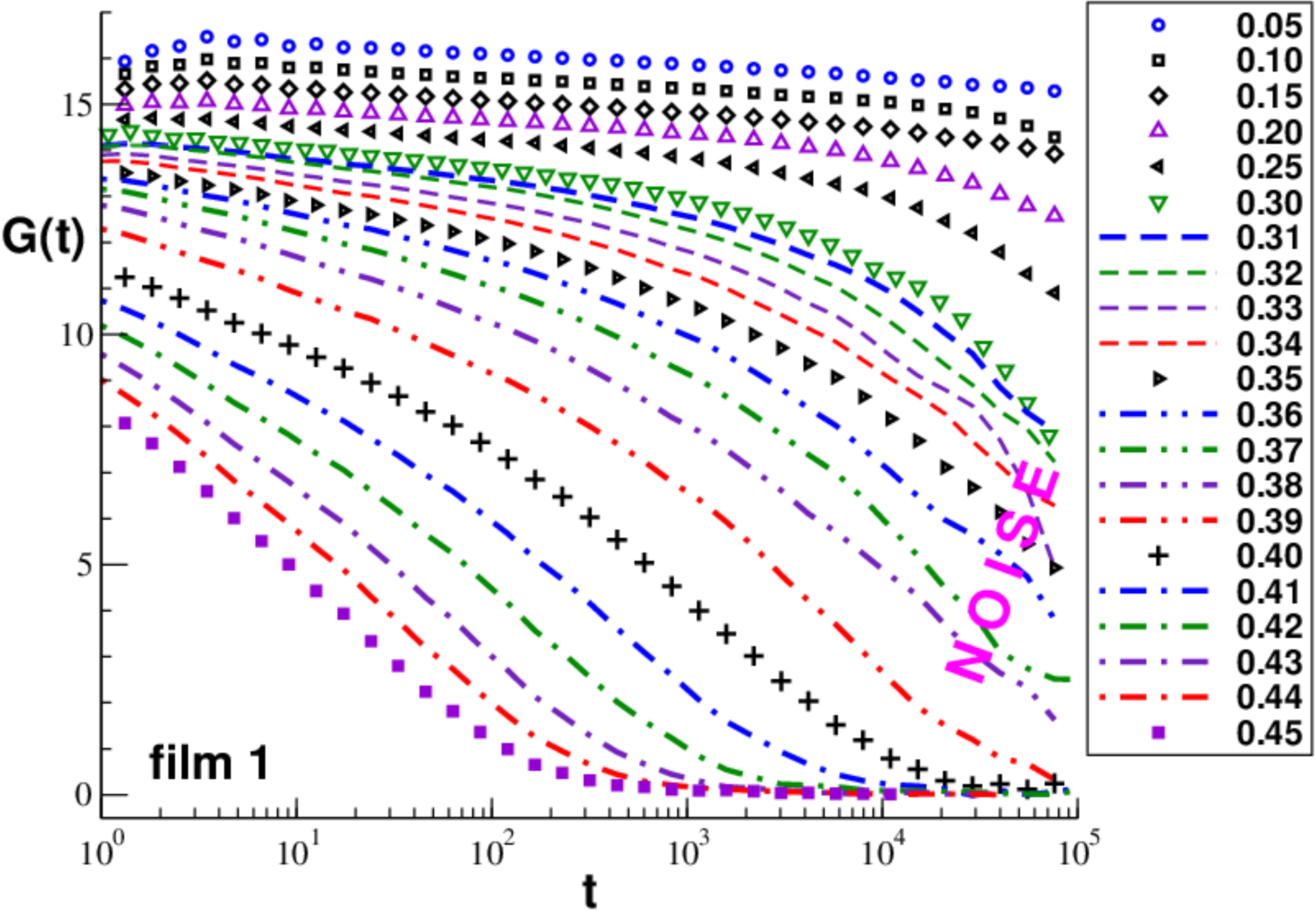}}}
\caption{Unscaled stress relaxation modulus $G(t)$ for film 1 using half-logarithmic coordinates. 
No indication of a jump singularity with respect to temperature is found.
}
\label{fig_Gt}
\end{figure}

\begin{figure}[t]
\centerline{\resizebox{1.0\columnwidth}{!}{\includegraphics*{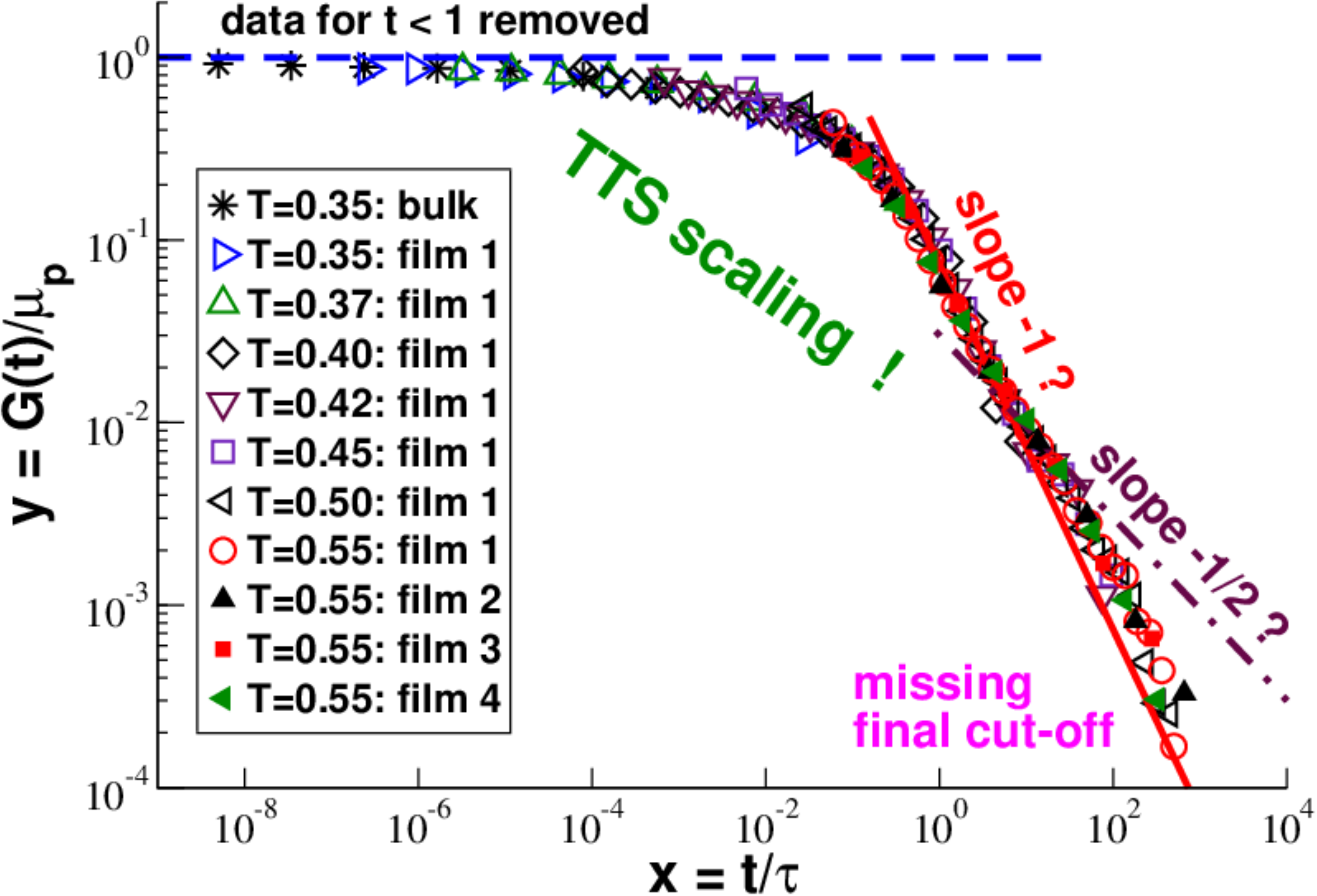}}}
\caption{Successful TTS scaling plot of $y=G(t)/\GFplat$ as a function of reduced
time $x = t/\tauterm$ using the same relaxation times as in Fig.~\ref{fig_GF_TTS}.
The two indicated power laws (bold and dash-dotted lines) are given for comparison.
Unfortunately, our production runs are too short to reveal the expected
final exponential cut-off even for the highest temperatures. 
}
\label{fig_Gt_TTS}
\end{figure}

Focusing on our thickest films and using a half-logarithmic representation, 
Fig.~\ref{fig_Gt} presents $G(t)$ for all temperatures $T \le 0.45$. 
Please note that albeit we ensemble-average over $m=120$ independent configurations 
it was necessary for the clarity of the presentation to use in addition gliding averages 
over the total production runs, i.e. the statistics becomes worse for $t \to \tsampmax=10^5$, 
and, in addition, to strongly bin the data logarithmically. 
Without this strong averaging the data would appear too noisy for temperatures around $\Tglass$.
(See Sec.~\ref{res_dGdmu} for a discussion of the standard deviation $\dGt$ of $G(t)$.)
However, it is clearly seen that $G(t)$ increases {\em continuously} with decreasing $T$ without 
any indication of the suggested jump-singularity 
\cite{Szamel11,Ikeda12,Yoshino14,Klix12,Klix15}.
This is consistent with the continuous decay of the storage modulus $G^{\prime}(\omega=const,T)$
as a function of temperature $T$ shown in Fig.~6 of Ref.~\cite{Pablo05b}. Similar continuous 
behavior has also been reported for the Young modulus of polymer films \cite{Riggleman13}.

Using a similar double-logarithmic representation as in Fig.~\ref{fig_GF_TTS},
we demonstrate in Fig.~\ref{fig_Gt_TTS} that a successful TTS scaling can 
be achieved for $G(t)$ just as for $\GF(\tsamp)$. 
While several temperatures are again indicated for film 1,
only one temperature is indicated for the other ensembles.
The effective power law $-1$ seen for $x \approx 1$ (solid line) can of course not 
correspond to the asymptotic long-time behavior since
\begin{equation}
\int_0^{\infty} \ddiff t \ G(t) = \eta
\mbox{ and }
\int_0^{\infty} \ddiff t \ t \ G(t) = \tauterm \eta
\label{eq_Gtint}
\end{equation}  
would diverge. We remind that the Rouse behavior expected to hold for our short chains
for large times corresponds to a cut-off with $y(x) \approx \exp(-x)/\sqrt{x}$ 
\cite{RubinsteinBook} for which all moments of $G(t)$ converge.
Basically, due to the not accessible final cut-off it is yet impossible for any 
temperature $T \le 0.55$ to determine $\eta$ and $\tauterm$ merely by integrating $G(t)$, Eq.~(\ref{eq_Gtint}),
and neither is it possible to compute $J(t)$ by Laplace transformation of $G(t)$ \cite{FerryBook} 
in order to compare our numerical results with recent experiments 
\cite{McKenna05,McKenna08,McKenna17,BodiguelFretigny:EPJE2006}.
It is mainly for this reason that we proceeded above by using the Einstein-Helfand 
relation and the TTS scaling of $\GF(\tsamp)$ to estimate $\eta$ and $\tauterm$.
The unfortunate intermediate effective power-law slope $-1$ observed in Fig.~\ref{fig_Gt_TTS} 
is presumably due to an intricate crossover between the exponential decay of the local glassy 
dynamics and the $1/\sqrt{x}$-decay (dash-dotted line) due to the chain connectivity. 
Albeit we do not expect any conceptional problems,
much longer production runs are clearly warranted to clarify this issue. 

\begin{figure}[t]
\centerline{\resizebox{1.0\columnwidth}{!}{\includegraphics*{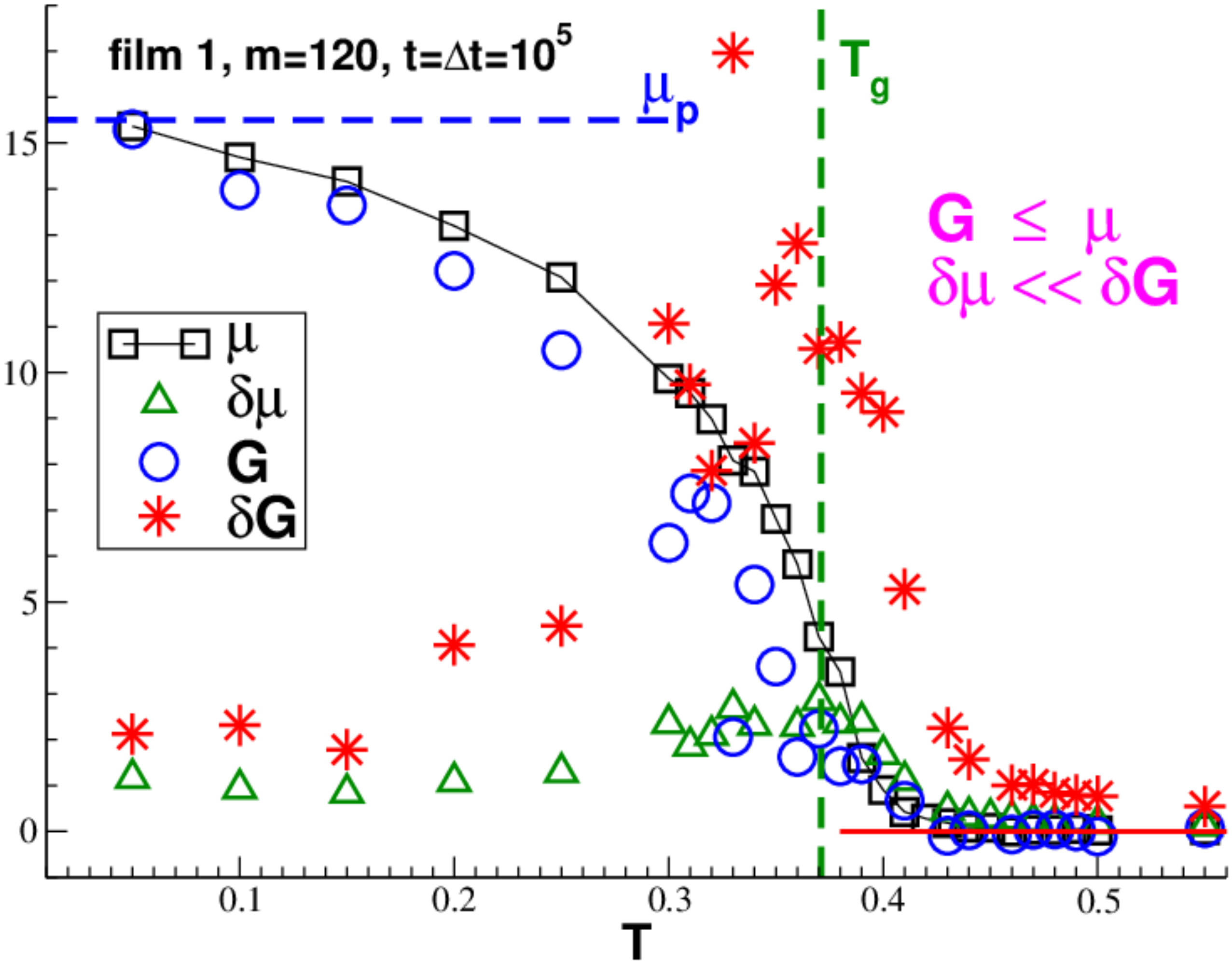}}}
\caption{Shear modulus $\GF$, shear relaxation modulus $G$
and the corresponding standard deviations $\dGF$ and $\delta G$
taken at $t=\tsamp=\tsampmax=10^5$ as functions of $T$. Focusing on film 1 all data are averaged 
over $m=120$ configurations without additional gliding averages and logarithmic binning.
The observed two inequalities $G \le \GF$ and $\delta G \gg \delta \GF$ are both
consequences of the stationarity relation Eq.~(\ref{eq_muGt}).
}
\label{fig_dGdmu}
\end{figure}

\subsection{Standard deviations $\dGF$ and $\delta G$}
\label{res_dGdmu}
As already pointed out above, the data for $G(t)$ is quite noisy, especially around $\Tglass$,
and we had to use gliding averages and a strong logarithmic binning for the clarity of the 
presentation.
We want now to describe this qualitative observation in more quantitative terms.
This is done in Fig.~\ref{fig_dGdmu} (focusing again on film 1) where we compare $\GF$ and $G$ 
and their respective standard deviations $\dGF$ and $\delta G$, Eq.~(\ref{eq_standderiv}),
taken at the same constant time $t=\tsamp=\tsampmax=10^5$ and plotted as functions of 
the temperature $T$. 
(The corresponding error bars $\delta \GF/\sqrt{m-1}$ and $\delta G/\sqrt{m-1}$ are not shown.)
While we still average over the $m$ independent configurations, we do not use any gliding 
averaging or logarithmic binning.

As already presented in Fig.~\ref{fig_key}, $\GF(T)$ decreases both continuously and smoothly with $T$. 
Albeit $G(T)$ decreases also continuously, it reveals an erratic behavior for temperatures slightly 
below $\Tglass$ (vertical dashed line).
The inequality $G(T) \le \GF(T)$ for all temperatures is expected from Eq.~(\ref{eq_inequal}). 
More importantly, being the second integral over $G(t)$, the shear modulus $\GF$ automatically 
filters off the high-frequency noise. This explains the observed strong inequality 
$\delta \GF \ll \delta G$ of the standard deviations.
At variance to $\GF$ and $G$, a striking non-monotonic behavior is observed for $\delta \GF$ and $\delta G$ 
with maxima slightly below the glass transition temperature $\Tglass$.
While $\delta \GF \ll \GF$ and $\delta G \ll G$ in the solid limit, $\delta \GF > \GF$ and $\delta G > G$
at high and intermediate temperatures. It can be demonstrated that $\delta \GF/\GF \approx \sqrt{2}$ 
holds in this limit (as expected for general Gaussian fluctuating fields). Unfortunately,
our statistics is insufficient to precisely quantify $\delta G$ or $\delta G/G$. 
However, it should be clear from the presented data that the glass transition is masked 
--- quite similar to what has been observed for 3D bulk systems \cite{ivan17c,ivan18a} ---
by strong ensemble fluctuations with $\delta \GF/\GF$ and $\delta G/G$ of order of unity.
The prediction of $G(T)$ or $\GF(T)$ for $T \approx \Tglass$ becomes thus meaningless 
for a single configuration.
We emphasize finally that the inequalities $\delta \GF \ll \delta G$
and $\delta \GF/\GF \ll \delta G/G$ are the strongest slightly below $\Tglass$. 
This is the main reason why a numerical study of the elastic shear strain response 
around the glass transition should better focus on $\GF$ rather than of $G$.

\section{Conclusion}
\label{sec_conc}
\paragraph*{Methodology.}
Free-standing polymer films (Fig.~\ref{fig_sketchSetup}) have been investigated by means of MD 
simulation of a standard coarse-grained polymer glass model (Appendix~\ref{app_Hamiltonian}).
The film thickness $H \sim 1/L^2$ was tuned by varying the lateral box width $L$.
The glass transition temperature $\Tglass$ was obtained from the much weaker temperature 
dependence of $H$ (Fig.~\ref{fig_H_T}).
We have focused on the global in-plane shear stresses (Appendix~\ref{app_tauAmuA}), their 
fluctuations (Sec.~\ref{res_SFF}) and relaxation dynamics (Figs.~\ref{fig_EH}-\ref{fig_Gt_TTS}). 
We used as the main diagnostic tool the first time-averaged and then ensemble-averaged
(Appendix~\ref{app_data}) shear modulus $\GF$ and its various contributions 
as defined by the stress-fluctuation formula, Eq.~(\ref{eq_SFF}). 

\paragraph*{$\tsamp$-dependence of $\GF$ and TTS scaling.}
As expected from previous work \cite{WXB16,ivan17c,ivan18a}, 
$\GF$ decreases monotonically (Figs.~\ref{fig_tsamp}-\ref{fig_GF_TTS}) with the 
sampling time $\tsamp$. 
This $\tsamp$-dependence is perfectly described (Fig.~\ref{fig_tsamp}) by the 
stationarity relation Eq.~(\ref{eq_muGt}), i.e. the stress-fluctuation formula 
is equivalent to a second integral over the shear stress relaxation modulus $G(t)$. 
The crucial consequence from the computational perspective is that, filtering away 
the high-frequency noise, $\GF(\tsamp)$ is a natural smoothing function
statistically much better behaved as $G(t)$.
As shown from the standard deviations $\delta \GF$ and $\delta G$ (Fig.~\ref{fig_dGdmu}), 
this is especially important for large times and for temperatures around the glass transition.
While the shear viscosities $\eta$ for the highest temperatures may be directly
computed by means of the Einstein-Helfand relation for $\GF(\tsamp)$, Eq.~(\ref{eq_EH}),
this is currently impossible using the corresponding Green-Kubo relation for $G(t)$, 
Eq.~(\ref{eq_Gtint}). 
Using the accurate TTS scaling of $\GF$ (Fig.~\ref{fig_GF_TTS}) we are able to estimate  
$\eta(T) \sim \tauterm(T)$ for an even broader temperature range down to $\approx \Tglass$. 
The TTS scaling of $G(t)$ is then possible (Fig.~\ref{fig_Gt_TTS}) using the same rescaling parameters.

\paragraph*{Continuous temperature behavior.}
In agreement with recent studies of 3D polymer glass-formers \cite{ivan17c,ivan18a}, 
$\GF$ and $G$ are found to decrease both monotonically and continuously with temperature $T$
(Figs.~\ref{fig_key},~\ref{fig_EH}-\ref{fig_dGdmu}). This result is qualitatively incompatible 
with mean-field theories \cite{Yoshino12,Yoshino14,Klix12,Klix15} which find that the energy 
barriers for structural relaxation diverge at the glass transition causing the sudden 
arrest of liquid-like flow. Non-mean-field effects smearing out the transition are apparently crucial.
The idea that correlations may matter around $\Tglass$ is strongly supported by the remarkable 
peaks observed for the standard deviations $\delta \GF$ and $\delta G$ (Fig.~\ref{fig_dGdmu}).

\paragraph*{Film thickness effects.}
As expected assuming a linear superposition of bulk and surface properties, Eq.~(\ref{eq_Acalsurfeff}), 
the glass transition temperature $\Tglass$ decreases linearly with $1/H$ (Fig.~\ref{fig_H_T}). 
Consistently, $\GF$ becomes finite at lower temperatures for thinner films 
(Fig.~\ref{fig_key}). The same linear superposition relation characterizes 
$\GF$ and its various contributions if taken in the low or high temperature limit 
(Figs.~\ref{fig_muA_T} and \ref{fig_muF_T}), the shear modulus $\GFglass$ at the glass transition 
and the plateau modulus $\GFplat$ (Fig.~\ref{fig_GF_Lscal}). 
Importantly, as shown in Fig.~\ref{fig_GF_TTS} and Fig.~\ref{fig_Gt_TTS},
it is possible to collapse $\GF(\tsamp)$ and $G(t)$ for all our ensembles 
using the strongly $H$-dependent relaxation time $\tauterm$.
(The weak $H$-dependence of the plateau modulus $\GFplat$ used for dimensionless reasons is 
less relevant for the scaling.)
Moreover, since $\tauterm(T,H)$ is found to roughly scale as a function of the inverse
reduced temperature $x=\Tglass(H)/T$ (Fig.~\ref{fig_GF_TTS}), the $H$-dependencies of 
all standard viscoelastic properties \cite{FerryBook} are essentially traced back to $\Tglass(H)$. 

\paragraph*{Discussion.}
While the shear viscosity and the terminal relaxation time at constant $T$ are linear in $1/H$ 
for high temperatures (inset of Fig.~\ref{fig_EH}) where $\tauterm$ is a weak function of $x$, 
for temperatures close to $\Tglass$ this can only be 
the leading contribution of a more general $1/H$-expansion. 
Due to the strong $x$-dependence of $\tauterm(x)$ for $x \approx 1$ (Fig.~\ref{fig_GF_TTS}), 
a weak variation of $1/H$ close to the glass transition must have a dramatic and in general 
non-linear effect on the thickness dependence of various viscoelastic properties. 
As already pointed out elsewhere \cite{Peter07},
some care is thus needed if $\Tglass(H)$ is operationally obtained by means of a rheological property 
other than Eq.~(\ref{eq_tauTg}). This may be an explanation for some of the $1/H$-expansions with higher 
order terms reported for $\Tglass(H)$ in the literature \cite{Vogt18,MangalaraEtal:JCP2017}.  

\paragraph*{Outlook.}
We are currently investigating the $z$-profiles of various properties considered here 
in order to confirm the superposition of bulk and surface properties and to demonstrate 
that Eq.~(\ref{eq_muGt}) also holds for $\GF(\tsamp,z)$ and $G(t,z)$.
The prefactor $c=1$ used for the terminal relaxation time $\tauterm$ was somewhat arbitrary, 
Eq.~(\ref{eq_taueta}). 
This was due to the missing exponential cut-off of $G(t)$ which made it impossible to determine $\tauterm$ 
accurately using Eq.~(\ref{eq_Gtint}) even for $T=0.55$. We plan to do this at least for one high temperature
using much longer production runs with $\tsampmax=10^7$. Using these longer time series it should be 
possible to fit the Maxwell relaxation spectrum \cite{FerryBook}. Together with an improved TTS scaling 
of $G(t,T)$ this should allow us to obtain $\GstorT$ and $\GlossT$ and to compare our data with the 
experimentally available creep compliance $J(t,T)$ 
\cite{McKenna05,McKenna08,McKenna17,BodiguelFretigny:EPJE2006}.
In addition we will attempt to characterize in more detail the scaling of the fluctuations between 
different configurations of the ensemble with the number of chains, the film volume,
the film thickness and the sampling time. A quantitative theoretical theory describing 
the standard deviations $\delta \GF$ and $\delta G$, especially around $\Tglass$, is highly warranted.

\begin{acknowledgments}
We are indebted to L.~Klochko and A.N. Semenov (both ICS, Strasbourg) for helpful discussions.
We thank the IRTG Soft Matter (Freiburg, Germany) for financial support and
the University of Strasbourg for CPU time through GENCI/EQUIP${@}$MESO.
\end{acknowledgments}

\appendix
\section{Instantaneous properties}
\label{app_useful}

\subsection{Canonical affine transform}
\label{app_afftrans}

Let us consider an infinitesimal simple shear strain increment $\gamma$ in the $xy$-plane 
as it would be used to determine the shear relaxation modulus $G(t)$ by means of a direct 
out-of-equilibrium simulation (Sec.~\ref{res_Gt}).
For simplicity all particles are in the principal simulation box \cite{AllenTildesleyBook}.
It is assumed \cite{WXP13} that all particle positions $\rvec$ and particle momenta $\pvec$
follow the imposed ``macroscopic" strain in a {\em canonical affine} manner according to 
\begin{equation}
\rx \to \rx + \gamma \ \ry \mbox{ and } 
\py \to \py - \gamma \ \px
\label{eq_cantrans}
\end{equation}
where the negative sign in the second transform assures that Liouville's theorem is satisfied.
Please note that a general configuration will (except for very simple lattice systems)
not follow an external macroscopic strain in an affine manner. The assumed transform is merely
a theoretical trick \cite{Lutsko88,WXP13}.

\subsection{Shear stress and affine shear modulus}
\label{app_tauAmuA}
The instantaneous shear stress $\tauhat$ and the instantaneous affine shear modulus $\muAhat$
are defined by the first two functional derivatives 
\cite{Lutsko88,WXP13,ivan18a}
\begin{equation}
\tauhat \equiv \left.\frac{\delta \ehat(\gamma)}{\delta \gamma}\right|_{\gamma=0}
\mbox{ and }
\muAhat 
\equiv  \left.\frac{\delta^2 \ehat(\gamma)}{\delta \gamma^2}\right|_{\gamma=0}
\label{eq_globalhat}
\end{equation}
of the energy density $\ehat \equiv \Ehat/V$ of the total energy $\Ehat$
with respect to a canonical affine transform defined above.
(We remind that for films $V=L^2H$ with $H$ being the film thickness defined in Sec.~\ref{res_density}.)
Assuming the energy $\Ehat = \Eidhat + \Eexhat$ to be the sum of an ideal and 
an excess contribution $\Eidhat$ and $\Eexhat$, similar relations apply for the corresponding 
contributions $\tauidhat$ and $\tauexhat$ to  $\tauhat =\tauidhat + \tauexhat$ and 
for the contributions $\muAidhat$ and $\muAexhat$ to $\muAhat = \muAidhat + \muAexhat$.
With $\Eidhat = \sum_{i=1}^{n} \pvec_i^2/2m$ being the standard kinetic energy for monodisperse
particles of mass $m$, Eq.~(\ref{eq_globalhat}) implies the ideal contributions
\begin{eqnarray}
\tauidhat & = & - \frac{1}{V} \sum_{i=1}^n \pix \piy / m \label{eq_tauidhat} \mbox{ 
and } \\
\muAidhat & = & \frac{1}{V} \sum_{i=1}^n (\pix^2 + \piy^2) /2m \label{eq_muidhat} 
\end{eqnarray}
where the sums run over all $n$ particles. Note that the minus sign for the ideal shear stress 
follows from the minus sign in Eq.~(\ref{eq_cantrans}) required for a canonical transform.
We have used a symmetric representation in Eq.~(\ref{eq_muidhat})
exchanging $x$ and $y$ in Eq.~(\ref{eq_cantrans}) and averaging over the equivalent
canonical affine simple shear strains in $x$ and $y$ directions.
Assuming a pairwise central conservative potential $\Eexhat = \sum_l u(\rl)$
with $l$ labeling the interactions, $\rl$ the distance between the pair of monomers
and $u(r)$ a pair potential as defined in Appendix~\ref{app_Hamiltonian},
one obtains the excess contributions 
\begin{eqnarray}
\tauexhat & = & \frac{1}{V} \sum_l \rl u^{\prime}(\rl) \ \nlx \nly   \label{eq_tauexhat} \ \mbox{ and } \\
\muAexhat & = & \frac{1}{V} \sum_l  \left( \rl^2 u^{\prime\prime}(\rl)
- \rl u^{\prime}(\rl) \right) \nlx^2 \nly^2 \nonumber \\
& + & \frac{1}{V} \sum_l \rl u^{\prime}(\rl) \ (\nlx^2 + \nly^2)/2  \label{eq_muAexhat}
\end{eqnarray}
with $\nvecl = \rvecl/\rl$ being the normalized distance vector.
As one expects, Eq.~(\ref{eq_tauexhat}) is strictly identical to the
corresponding off-diagonal term of the Irving-Kirkwood stress tensor  \cite{AllenTildesleyBook}.
We have again used a symmetric representation for the last term in Eq.~(\ref{eq_muAexhat}).
Importantly, this term takes into account the excess contribution of the normal tangential stresses 
in the $(x,y)$-plane. These contributions cannot be neglected for stable films with finite surface tension. 
This last term corresponds to the well-known Birch coefficients \cite{SBM11,FrenkelSmitBook} 
contributing to the elastic moduli of stressed systems.
We also note that $\muAexhat$ depends on the second derivative $u^{\prime\prime}(r)$ of the pair potential. 
Impulsive corrections need to be taken into account due to this term if the first derivative $u^{\prime}(r)$ 
of the potential is not continuous \cite{XWP12}. Unfortunately, this is the case at the cut-off of the 
shifted LJ potential, Eq.~(\ref{eq_ULJshift}), used in the current study.

\section{Computational details}
\label{app_algo}

\subsection{Model Hamiltonian}
\label{app_Hamiltonian}
All monomers, that are not connected by bonds, interact basically via a monodisperse LJ potential 
\cite{AllenTildesleyBook}
\begin{equation}
\ULJ(r) = \epsLJ \left( (\sigLJ/r)^{12} - (\sigLJ/r)^6 \right).  
\label{eq_ULJ}
\end{equation}
LJ units \cite{AllenTildesleyBook} are used throughout this work,
i.e. the monomer mass $m$, the monomer diameter $\sigLJ$, the LJ energy parameter $\epsLJ$
and Boltzmann's constant $\kB$ are all set to unity. Length scales are given 
in units of $\sigLJ$, energies in units of $\epsLJ$,
stresses and elastic moduli in units of $\epsLJ/\sigLJ^3$
and times in units of $\sqrt{m\sigLJ^2/\epsLJ}$.
The LJ potential is truncated at $\rcut=2.3 \approx 2 \rmin$, with $\rmin=2^{1/6}$ 
being the potential minimum, and shifted 
\begin{equation}
\ULJshift(r) = \ULJ(r) - \ULJ(\rcut) \mbox{ for } r \le \rcut
\label{eq_ULJshift}
\end{equation}
to make it continuous.
It is, however, not continuous with respect to its first derivative and impulsive 
truncation corrections \cite{FrenkelSmitBook} are thus required for the determination 
of the Born-Lam\'e coefficients \cite{XWP12,WXP13}.
The flexible bonds are represented by the spring potential
\begin{equation}
\Ubond(r) = \frac{\kbond}{2} \ (r- \lbond)^2
\label{eq_Ubond}
\end{equation}
with $r$ being the distance between the permanently connected beads, 
$\kbond = 1110$ the spring constant and $\lbond=0.967$ the bond length.

\subsection{Data sampling and averaging procedures}
\label{app_data}

Instantaneous observables $\ahat$ are sampled every $10 \dtMD$ with $\dtMD=0.005$ 
being the time increment of the velocity-Verlet scheme used.
Of central importance are the instantaneous shear stress $\tauhat$ and the instantaneous 
affine shear modulus $\muAhat$ defined in Appendix~\ref{app_tauAmuA}.
Note that all intensive properties are normalized using the effective film volume $V=L^2H$ 
with $H$ being the film thickness defined in Sec.~\ref{res_density}.
As described in detail in Ref.~\cite{ivan18a}, the stored time-series are used to compute for a 
given configuration various (arithmetic) {\em time averages} (marked by horizontal bars) 
\begin{equation}
\overline{\ahat} \equiv \frac{1}{i_2-i_1+1} \sum_{i=i_1}^{i_2} \ahat_i
\label{eq_timeaver}
\end{equation}
with $i = t/(10 \dtMD)$ being the index of the time series
and the sum running over all data entries of the time window $(t_1,t_2=t_1+\tsamp)$
with $\tsamp \le \tsampmax=10^5$ being the sampling time.
By averaging over the $m$ independent configurations,
we obtain then {\em ensemble averages} (marked by pointy brackets)
\begin{equation}
\la \Acal_j \ra \equiv \frac{1}{m} \sum_{j=1}^m \ \Acal_j 
\label{eq_ensaver}
\end{equation}
with $j$ being the configuration index and $\Acal_j$ some function of time preaveraged properties. 
The standard deviations $\delta \GF$ and $\delta G$ discussed in Sec.~\ref{res_dGdmu} are obtained using
\begin{equation}
\sqrt{\la \Acal_j^2 \ra -\la \Acal_j \ra^2} \mbox{ with }
\Acal_j = \overline{\GF} \mbox{ or } = \overline{G} 
\label{eq_standderiv}
\end{equation}
being, respectively, the shear modulus or the relaxation modulus for a given time-window of a configuration.
Essentially, the same data averaging procedure is used for the bulk systems
the only difference being that we average finally in addition over 
the three equivalent shear planes. 

\subsection{Simple averages and fluctuations}
\label{app_simplefluctu}

It is important to distinguish in a computation study
between {\em ``simple averages"} $\langle \overline{\ahat} \rangle$
and {\em ``fluctuations"} such as
\cite{AllenTildesleyBook,WXB16,ivan18a} 
\begin{equation}
\la \overline{\ahat}^2 \ra \mbox{ or }
\la \overline{(\ahat-\overline{\ahat})}^2 \ra 
= \la \overline{ \ahat^2} - \overline{\ahat}^2\ra.
\label{eq_fluctu}
\end{equation}
It is well known that simple averages and fluctuations behave differently
under ensemble transformation \cite{AllenTildesleyBook,WXB16}. 
Incidentally, using the Lebowitz-Percus-Verlet transformation rules 
this provides one way to elegantly demonstrate the stress-fluctuation 
formula Eq.~(\ref{eq_SFF}) within a couple of lines \cite{WXP13,WXB16}.
Interestingly, the expectation values, i.e. the ensemble averages for large $m$, 
of simple averages do not depend on the sampling time $\tsamp$
since their time and ensemble averages commute \cite{WXB16,ivan18a}
\begin{equation}
\la \overline{\ahat} \ra = \overline{ \la \ahat \ra } \sim \tsamp^0
\mbox{ since } \la \ahat \ra  \sim \tsamp^0.
\label{eq_commute}
\end{equation}
As emphasized in Sec.~\ref{res_tsamp}, this does not hold in general for fluctuations.
As reminded in Appendix~\ref{app_fluctu}, it is always possible for stationary systems to 
describe the $\tsamp$-dependence of time-preaveraged fluctuations in terms of a weighted integral 
over a corresponding correlation function.
The specific relation relevant for the present work is given by Eq.~(\ref{eq_muGt}).

\section{Fluctuations in stationary time series}
\label{app_fluctu}

\paragraph*{Time-translational invariance.}
Let us consider a time series $(x_1,\ldots,x_n, \ldots x_N)$  with entries $x_n$
sampled at equidistant time intervals $\ddiff t$.
The time-averaged variance of this time series may be rewritten as
\begin{eqnarray}
\overline{x^2}-\overline{x}^2 & = & \overline{(x_n-\overline{x})^2}=
\frac{1}{2N^2} \sum_{n,m=1} (x_n-x_m)^2 \nonumber\\
&=& \frac{2}{N^2} \sum_{s=0}^{N-1} (N-s) \ \overline{h}(s,N)
\label{eq_var_hs} 
\end{eqnarray}
using the in general $s$- and $N$-dependent sum
\begin{equation}
\overline{h}(s,N) \equiv \frac{1}{2} \ \frac{1}{N-s} \sum_{n=1}^{N-s} \ (x_{n+s}-x_n)^2.
\label{eq_hs_def}
\end{equation}
If time-translational invariance can be assumed on average, we can readily take the expectation value 
$\langle \ldots \rangle$ over an ensemble of such time series. This yields
\begin{eqnarray}
\la \overline{x^2}-\overline{x}^2 \ra & = & \frac{2}{N^2} \sum_{s=0}^{N-1} (N-s) \ h(s) 
\mbox{ with }
\label{eq_var_hsav} \\
h(s) & \equiv & \la \overline{h}(s,N) \ra = c(0)-c(t) \mbox{ and } 
\label{eq_hs_defav} \\
c(s) & \equiv & \la \overline{x_s x_0} \ra.
\label{eq_cs_defav}
\end{eqnarray}
Note that the mean-square displacement $h(s)$ and the correlation function $c(s)$
do only depend on the time-increment $s$ for stationary time series.

\paragraph*{Continuum limit.}
Using that the time series have been sampled with equidistant time steps, 
i.e. $t \approx s \ddiff t$ and $\tsamp \approx N \ddiff t$,
the latter result becomes in the continuum limit 
\begin{equation}
\la \overline{x^2}-\overline{x}^2 \ra = \Tver[h(t)] = c(0)-\Tver[c(t)] 
\label{eq_var_hscont}
\end{equation}
where we have used the useful linear functional
\begin{eqnarray} 
\Tver[y(t)] & \equiv& \frac{2}{\tsamp^2} \int_0^{\tsamp} \ddiff t \ (\tsamp-t) \ y(t) 
\label{eq_Tver_def} \\
            & =     & \frac{2}{\tsamp^2} \int_0^{\tsamp}\ddiff t \int_0^t \ddiff \tp \ y(\tp).
\label{eq_Tver_def_two}
\end{eqnarray}
Note that contributions at the lower boundary of the integral have a strong weight 
due to the $(\tsamp-t)$-factor in Eq.~(\ref{eq_Tver_def}). 
If $c(t)$ is a strictly monotonically decreasing function, this implies the inequality 
\begin{equation}
c(t=\tsamp) < \Tver[c(t)].
\label{eq_inequal}
\end{equation}

\paragraph*{Back to current problem.}
Setting $x(t) \equiv \sqrt{\beta V} \tauhat(t)$ and assuming time translational invariance for
the sampled instantaneous shear stresses $\tauhat$, Eq.~(\ref{eq_var_hscont}) and Eq.~(\ref{eq_ht}) 
lead to 
\begin{equation}
\muF(\tsamp) \equiv \muFtwo - \muFone(\tsamp) = \Tver[h(t)]
\label{eq_muFtsamp}
\end{equation}
for the $\tsamp$-dependence of the shear-stress fluctuations.
Since $\muA$ is a constant, Eq.~(\ref{eq_Gtgeneral}) implies
\begin{eqnarray}
\GF(\tsamp) & \equiv & \muA - \muF(\tsamp) = \Tver[G(t)] \nonumber \\
            & = & \frac{2}{\tsamp^2} \int_0^{\tsamp}\ddiff t \int_0^t \ddiff \tp \ G(\tp) 
\label{eq_GFtsamp}
\end{eqnarray}
in agreement with Eq.~(\ref{eq_muGt}) stated in the Introduction.
If $G(t)$ approaches a final constant $\Geq$, as sketched in panel (b) of Fig.~\ref{fig_sketchSFF}, 
or a broad intermediate plateau, $\GF(\tsamp)$ must ultimately follow, however, more slowly being 
dominated by the short-time behavior of $G(t)$. 
We note finally that we might have also used $\GF(\tsamp) \equiv \Tver[G(t)]$ 
as the fundamental definition rather then the thermodynamically motivated 
stress-fluctuation formula, Eq.~(\ref{eq_SFF}).

\clearpage
\newpage


\begin{thebibliography}{71}
\bibitem[{foo({\natexlab{a}})}]{foot_samplingtime}
\bibinfo{note}{The sampling time $\tsamp$ is the time window over which
  observables are first time-averaged according to Eq.~(\ref{eq_timeaver}) for
  a given configuration before an ensemble average over $m$ independent
  configurations is performed. $\tsamp$ should not be confused with the waiting
  time for systems with strong aging effects \cite{ivan18a}.}
\bibitem[{foo({\natexlab{c}})}]{foot_muAz}
\bibinfo{note}{It can be shown that the $z$-distribution $\muA(z)$ has a broad
  surface regime at high temperatures explaining thus the strong
  $1/H$-corrections, while it vanishes suddenly at the surfaces at low
  temperatures.}
\bibitem[{foo({\natexlab{d}})}]{foot_muFsign}
\bibinfo{note}{The different signs of the correction can be explained from the
  qualitative different shapes of the $z$-distribution $\muF(z)$ in both
  temperature limits. While $\muF(z)$ is monomodal at large $T$ with a maximum
  at the film midplane, it becomes bimodal below $\Tglass$ with a minimum at
  the film midplane and maxima close to the surfaces. The reason for this is
  basically that the monomer mobility remains high at the surfaces, i.e.
  $\muF(z)$ can approach $\muA(z)$ from below, while $\muF(z) \ll \muA(z)$ in
  the bulk phase. Details will be given elsewhere.}
\end{thebibliography}

\end{document}